\documentclass[%
 reprint,
superscriptaddress,
 amsmath,amssymb,
 aps,
prl,
]{revtex4-2}
\usepackage{graphicx}
\usepackage{dcolumn}
\usepackage{bm}
\usepackage[colorlinks=true]{hyperref}  
\hypersetup{
    unicode=false,          
    pdftoolbar=true,        
    pdfmenubar=true,        
    pdffitwindow=false,     
    pdfstartview={FitH},    
    pdftitle={Flstar},    
    pdfauthor={Christos, Sachdev, Luo},     
    pdfsubject={},   
    pdfcreator={},   
    pdfproducer={}, 
    pdfkeywords={} {} {}, 
    pdfnewwindow=true,      
    colorlinks=true,       
    linkcolor=blue, 
    citecolor=blue,        
    filecolor=magenta,      
    urlcolor=blue           
} 
\usepackage{xcolor}
\newcommand{\Tr}{ \, \mathrm{Tr} \, }
\newcommand{\p}{\partial}
\newcommand{\eps}{\varepsilon}
\newcommand{\sgn}{\, \mathrm{sgn} \, }

\renewcommand{\vec}{\mathbf}
\renewcommand{\Re}{\,\mathrm{Re}\,}
\renewcommand{\Im}{\, \mathrm{Im}\,}

\newcommand{\pv}[1]{\textcolor{black}{#1}}
\newcommand{\sk}[1]{\textcolor{black}{#1}}

\usepackage{tikz-feynman}
\usepackage{pdfpages}
\makeatletter
\AtBeginDocument{\let\LS@rot\@undefined}
\makeatother

\begin{document}

\preprint{APS/123-QED}

\title{Strong non-linear response of strange metals}

\author{Serhii Kryhin}
\email{skryhin@g.harvard.edu }
\affiliation{Department of Physics, Harvard University, Cambridge MA 02138, USA}
\author{Subir Sachdev}
\email{sachdev@g.harvard.edu}
\affiliation{Department of Physics, Harvard University, Cambridge MA 02138, USA}
\author{Pavel A. Volkov}%
 \email{pavel.volkov@uconn.edu}
\affiliation{Deparment of Physics, University of Connecticut, Storrs, CT 06269, USA
}%

\date{\today}

\begin{abstract}
We show that nonlinear transport responses in strange metals are strong, larger by a factor of $E_F/T$ than in Fermi liquids. Within the two-dimensional Yukawa-Sachdev-Ye-Kitaev model of a Fermi surface with a spatially random coupling to a critical scalar, the third order conductivity is found to diverge as $1/T$ at low $T$, indicating the existence of a voltage-temperature scaling regime in the conductance. Its frequency and orientation dependence contains information on relaxation times of heat and electron distribution deformations, providing a new set of tools to characterize strange metals.
\end{abstract}

\maketitle


\emph{Introduction.}
The strange metal state remains one of the most enigmatic phenomena in correlated electron systems. Both its microscopic origin and definitive set of characteristic behaviors remain under debate, calling for new probes and predictions. Recent advances in THz optics have opened the way to probe nonlinear transport properties of correlated electronic systems at frequencies relevant for the low-energy electronic phenomena. So far, these techniques have found applications in probing collective modes in superconductors \cite{matsunaga2014light,higgs_rev,Wang_nonlin_22,salvador2024principles}, quantum spin systems \cite{armitage_qsl,takayoshi2019,parameswaran2020,choi2020} and strongly disordered semiconductors \cite{mahmood2021observation}. Metals remain relatively unexplored in this regard. Works on nonlinear optical conductivity in semiconductors \cite{Rustagi70,Rustagi84,Wolff82} have identified that nonparabolicity of the  band structure or energy dependence of scattering are necessary for any nonlinear response to be present (vanishing in Galilean invariant systems). In Fermi liquids, both effects are suppressed by the large value of the Fermi energy scale $T_F$, and thus the nonlinear response is expected to be weak.

\begin{figure}[t]
    \centering
    \includegraphics[width=1.0\columnwidth]{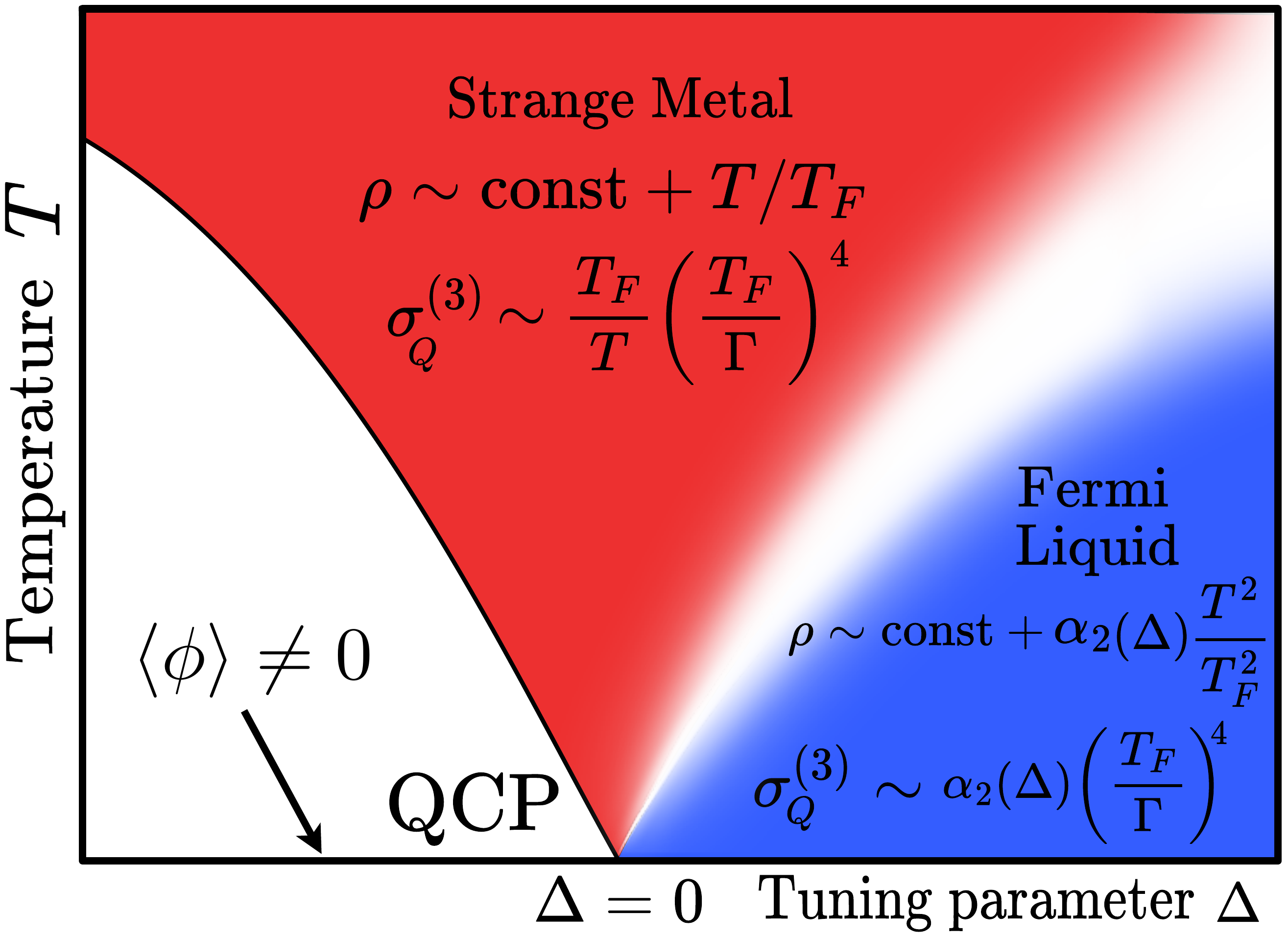}
    \caption{Behavior of linear and non-linear conductivity around a quantum critical point (QCP). As $\Delta$ is tuned to criticality, $\alpha_2(\Delta)$ grows until it saturates at $\sim T_F/T$ at the critical point. This leads to a crossover from a Fermi liquid (blue) into strange metal (red) both in linear and third order conductivities. The linear  conductivity $\sigma$ is in units of $e^2/ \hbar$, the third order conductivity $\sigma^{(3)}_Q$ is in units of $e^2/ \hbar E_0^2$, where $E_0 = 2 k_F^2 v_F/e$ is a characteristic electric field strength. 
    The state with $\langle \phi \rangle \neq 0$ either has a broken symmetry (which could extend to $T>0$), or is fractionalized if $\phi$ is a Higgs boson coupled to an emergent gauge field.}
    \label{fig:Fan}
\end{figure}

In this work we demonstrate that strange metals, in contrast, should exhibit strong nonlinear transport responses. Using the recently proposed two-dimensional Yukawa-Sachdev-Ye-Kitaev (2d-YSYK) model \cite{Aavishkar2023,PhysRevLett.133.186502}, we derive kinetic equations for non-equilibrium distributions of fermions and quantum critical bosons, allowing to treat the problem in presence of strong electric fields. The third order conductivity is found to be enhanced by $T_F/T$ with respect to the Fermi liquid state (see Fig. \ref{fig:Fan}) and a potential $E/T$ scaling regime (where $E$ is the electric field amplitude) in nonlinear transport is predicted to arise at low temperatures. We demonstrate that our results extend to a broad class of models realizing the marginal Fermi liquid phenomonology \cite{Varma1989} and give general arguments, relating strong nonlinear responses and linear in $T$ resistivity. These establish strong nonlinear response as a generic property of non-Fermi liquids.


\emph{Model.} To describe the strange metal, we use the 2d-YSYK model \cite{Altman1,Guo2021,Guo2022,Aavishkar2023,GuoIII,Nikolaenko2023,PhysRevLett.133.186502}, which has been shown to reproduce the characteristic linear transport ($\rho\sim T$ down to zero) and thermodynamic $C\sim T\log[1/T]$ properties of strange metals. It contains $N$ flavors of fermions $\psi_i$ with an action $S_\psi$, 
\begin{equation}
    S_\psi = \sum_{i = 1}^N \int dt d\vec r \; \psi_i^\dagger \left[ i \p_t - \eps(\hat {\vec p}) + \mu \right] \psi_i - \vec j_i \cdot \vec A, 
\end{equation}
where \pv{$j_i= \frac{e}{2im} (\hat \psi_{i+}^\dagger \nabla \hat \psi_{i+} - \mathrm{h.c.})$} 
, ${\bf A}$  is the vector potential, and $\varepsilon(p) \approx v_F(p-p_F)$ is the single-particle dispersion; and $N$ flavors of scalar bosons described by:
\begin{equation}
    S_\phi = \frac{1}{2} \sum_{i = 1}^N \int dt d\vec r \left[ \dot \phi_i^2 - c^2 (\nabla \phi_i)^2 - m^2 \phi_i^2 \right]
    \label{eq:sphi}
\end{equation}
in two spatial dimensions. We assume the system is at the QCP, such that the mass of the boson $m$ is tuned to 0 at zero temperature \cite{Guo2021}.
The final two ingredients are potential disorder for fermions, and the spatially disordered Yukawa coupling between fermions and bosons. We do not consider the effects of a uniform Yukawa coupling \cite{Guo2022} (which cancels in perturbative computations of transport \cite{Aavishkar2023}) or random mass disorder for bosons (which is induced by the Yukawa coupling, and is only important at very low $T$ \cite{lunts2024}). The action for fermion potential disorder $S_v$ has the form
\begin{equation}
S_v = - \sum_{i,j}\int dt d\vec r \; \frac{v_{ij}(\vec r)}{\sqrt{N}} \psi_i^\dagger(t, \vec r) \psi_j(t, \vec r)    
\end{equation}
with $v_{ij}(\vec r) = v_{ji}^*(\vec r)$
and the fermion-boson interaction comes in the form of $S_{g^\prime}$, where
\begin{equation}
    S_{g^\prime} = - \sum_{ijl}\int dt d\vec r \; \frac{c g^\prime_{ijl}(\vec r)}{N} \psi_i^\dagger(t, \vec r) \psi_j(t, \vec r) \phi_l(t, \vec r)
\end{equation}
with $g^\prime_{ijl}(\vec r) = g^\prime_{jil}(\vec r)$.
Disorder strength $v_{ij}(\vec r)$ and coupling strength $g^\prime_{ijl}(\vec r)$ are random functions of space and particle flavor such that
\begin{align}
    \langle v_{ij}(\vec r) v_{nm}^*(\vec r^\prime) \rangle &= v^2 \delta_{in} \delta_{j m} \delta(\vec r - \vec r^\prime),
    \\
    \langle g_{ijl}(\vec r) g_{nms}^{\prime *} (\vec r^\prime) \rangle &= g^{\prime 2} \delta_{in} \delta_{jm} \delta_{ls} \delta(\vec r - \vec r^\prime).
\end{align}
The total action of the model $S$ then can be written as
\begin{equation}
    S[\psi, \psi^\dagger, \phi] = S_\psi + S_\phi + S_v + S_{g^\prime}.
\end{equation}

Previously, kinetic equations for similar models have been derived \cite{arovas2018,Nikolaenko2023} assuming the bosons to be in thermal equilibrium. As is shown below, this is sufficient to describe the linear response, since the bosons don't directly couple to electric field ${\bf E}$. In higher orders in ${\bf E}$, however, the boson distribution also changes, affecting the electronic responses. We address this challenge by deriving a self-consistent set of effective kinetic equations for the fermions and bosons in the Yukawa-SYK model using the Keldysh formalism in the closed time contour formulation \cite{Kamenev_2011} with ``$+$" part going from $-\infty$ to $+ \infty$ and ``$-$" part vice versa.

Physical observables in the non-equilibrium field theory can be conveniently expressed via correlators of ``$+$" and ``$-$" fields. For example, the physical current operator $
    \hat{\vec j} \equiv \frac{e}{2imN}\sum_i \left[\hat \psi_{i+}^\dagger \nabla \hat \psi_{i+} - \nabla \hat \psi_{i+}^\dagger \hat \psi_{i+}\right]$ \pv{\footnote{We ignore spin, which amounts to a factor of 2.}}
can be expressed through a `lesser' Green's function $G^<(x, x^\prime) = \langle \psi_+(x) \psi_-^\dagger(x^\prime) \rangle$ (where $x=({\bf r},t)$)
as
\begin{equation}
    \langle \hat j(t, \vec r) \rangle = - \frac{e}{2 m} \lim_{\substack{t^\prime \rightarrow t+0 \\ \vec r^\prime \rightarrow \vec r}} [\nabla_{\vec r} G^<(x, x^\prime) - \nabla_{\vec r^\prime} G^<(x, x^\prime)],
    \label{eq:curdef}
\end{equation}
since $\langle \psi_+(x) \psi_+^\dagger(x^\prime) \rangle = G^<(x, x^\prime)$ for $t^\prime > t$.

As we are primarily interested in the low energy/wavelength behavior of the theory, we work with a Wigner transform - a Fourier transform of $G^<(x, x^\prime)$ around the center of mass coordinate $(x+ x^\prime)/2$ that we denote as $G^<((x+x^\prime)/2, \omega, \vec k)$. We focus on quasi-classical sources, with a uniform electric field with sufficiently small 
frequency ($\nu \ll T$, see below). In equilibrium $G^<(\omega, \vec k) = -2 i \bar f(\omega) \Im G_R(\omega, \vec k)$, where $\bar f(\omega)$ is the Fermi-Dirac distribution and $G_R(\omega, \vec k)$ is the retarded Green's function \cite{Nave2007,Nikolaenko2023}. An analogous statement is also true for the bosonic ``lesser" Green's function $D^<$ \cite{Supplement}. Motivated by these properties, we define the non-equilibrium ``occupation number function" $f(x, \omega, \theta)$ as
\begin{equation}
    f(x, \omega, \theta) = \frac{i}{2 B_F} \int_{- \infty}^{+\infty} \frac{dk}{(2 \pi)} G^{<}(x, \omega, \vec k),
    \label{eq:fdef}
\end{equation}
where $B_F = \int_{-\infty}^{+\infty} \pv{\frac{dk}{2\pi}} \, \Im G_R(x, \omega, \vec k) \pv{= - \frac{1}{2v_F}}$. The angle $\theta$ defines direction of $\vec k$ in the spatial plane. If the Fermi velocity of the system is large, the Green's function $G_R$ is highly peaked at $k=k_f$, making $B_F$ a constant up to $\omega/E_F$ corrections  \cite{Supplement}. In equilibrium, $f(x,\omega,\theta) = \bar f(\omega)$. 

An equivalent construction is used to define the non-equilibrium ``boson distribution" 
\begin{equation}
    f_B(x, \Omega, \theta) = -\frac{i}{2 B(\Omega)} \int_0^{+\infty} \frac{c^2 q dq}{2\pi} \, D^< (x, \Omega, \vec q),
    \label{eq:fbdef}
\end{equation}
where $\Omega$ is the boson energy and $\theta$ defines the direction of boson momentum, and $B = \int_0^\infty c^2 q dq / 2 \pi \, D_R(x, \Omega, \vec q)$. Unlike the fermion case, quantity $B$ ends up having non-trivial $\Omega$ dependence.

\emph{Kinetic equation.} To derive the set of equations closed for $f$ and $f_B$, we use the $\Sigma$-$G$ effective action method \cite{Guo2021,Guo2022} to obtain the equations for Green's functions in the large-$N$ limit, followed by a quasi-classical approximation developed in \cite{Nave2007}. As neither bosonic or fermionic self-energies depend on absolute value of momentum \cite{Aavishkar2023, Guo2021, Supplement}, the equations can be integrated over $k$, yielding closed system for $f$ \eqref{eq:fdef} and $f_B$ \eqref{eq:fbdef}.

The equations of motion can be simplified using angular harmonics $f_m$ and $f_{Bm}$, $f = \sum_m f_m e^{i \theta m}$ and $\quad f_B = \sum_m f_{Bm} e^{i \theta m}$. For $m \neq 0$ the result is (details in the End Matter and supplement \cite{Supplement}):
\begin{multline}\label{eq:fm}
    [a(T) \p_t + \Gamma_m + g(\omega, T) + \delta g[f_{B0}, f_0] ] f_{m\neq 0}
    = 
    \\
    = \pv{-} \frac{e v_F}{2} \, \p_\omega (\mathcal{E}^* f_{m-1} + \mathcal{E} f_{m+1}),
\end{multline}
where $\mathcal{E} = E_x + i E_y$. $\Gamma_{m\neq 0}= v^2 k_F/v_F$ is the relaxation rate associated with potential disorder. Below we will also discuss a more general case of $\Gamma_m$ being not all equal, expected when the discrete lattice symmetry is taken into account. $g(\omega, T)$ corresponds to the relaxation rate due to interactions between electrons and bosons, when both are taken close to equilibrium. In the vicinity of the QCP ($m^2$, Eq. \eqref{eq:sphi}, tuned to zero), $g \equiv {\rm Im} \Sigma_R =g_\mathrm{cr} = \alpha_1 T (\gamma(T) + \ln \mathrm{ch} (\omega / 2 T) )$, where $\alpha_1= \frac{g^{\prime 2} k_F}{4 \pi v_F}$ is the dimensionless coupling constant and $\gamma(T) = \ln (2/ \pi) + \ln \ln (\Lambda_q^2 c^2/c_d T)$, $c \Lambda_q \sim T_F$ is almost constant apart from extremely low $T$ (consistent with previous results \cite{Aavishkar2023,Guo2021,Guo2022,Nikolaenko2023} and marginal Fermi liquid phenomenology \cite{Varma1989}). In particular, at $\omega\gg T$, $g\sim |\omega|$ is a nonanalytic function.  In addition,  $a_\mathrm{cr}(T) = 1 + \alpha_1 \ln (\Lambda_q^2 c^2/2 \pi c_d T)/2\pi$ also shows a log divergence at the QCP. Away from the critical point $m^2=\Delta^2$, Fermi liquid behavior is obtained: $g_\mathrm{Fl} = 3\alpha_2(T^2 + \omega^2 / \pi^2)/4k_Fv_F$ with $\alpha_2 = g^{\prime 2} k_F^2 c_d/6 \Delta^2$, and the dynamic coefficient $a_\mathrm{FL}$ is constant at low $T$.

The non-equilibrium correction to the scattering rate $\delta g$ depends only on $m=0$ harmonics of $f,f_B$ and takes the form \cite{Supplement}
\begin{equation}
        \delta g [f_{B0}, f_0] = \frac{2 g^{\prime 2} k_F}{v_F} \int \frac{d \Omega}{2 \pi}  (K_{\delta g}[f_{B0}, f_0] - K_{\delta g}[\bar f_B, \bar f])
        \label{eq:deltag_main}
\end{equation}
with $K_{\delta g}[f_{B0}, f_0] = B(\Omega) (f_{B0}(\Omega) - f_0(\omega - \Omega))$. The kinetic equation for $f_0$ is more complicated than for $m\neq 0$ 
\begin{equation}\label{eq:f0}
        a(T) \p_t f_0 - I[f_{B0}, f_0,] = \pv{-}\p_\omega (\mathcal{E}^* f_{-1} + \mathcal{E} f_{+1})
\end{equation}
where
\begin{multline}\label{eq:I_f0}
    I[f_{B0}, f_0] = -  g(\omega, T) (f_0 - \bar f) - (f_0 - 1/2) \delta g[f_{B0}, f_0] +
    \\
    + \frac{g^{\prime 2} k_F}{2 v_F} \int \frac{d\Omega}{2 \pi}  (K_{g^\prime}[f_{B0}, f_0] - K_{g^\prime}[\bar f_B, \bar f]),
\end{multline}
and $K_{g^\prime}[f_{B0}, f_0] = B(\Omega) (2 f_{B0}(\Omega) + 1) (f_0(\omega + \Omega) + f_0(\omega - \Omega) - 1)$. Dynamics of the $m=0$ harmonic involves the deviations from equilibrium of both fermions and bosons in an essential way and has to be included to obtain nonlinear responses. Note the absence of contribution of potential scattering to \eqref{eq:f0}; this is due the $m=0$ harmonic characterizing the change in energy of the system, so elastic scattering can not lead to its relaxation. For a closed system of fermions and bosons, one furthermore expects energy conservation to impose zero energy relaxation rate, as shown below. 

The dynamics of boson distribution is dominated by the Landau damping such that the steady state $f_B$ can be found explicitly \cite{Supplement}:
\begin{equation}\label{eq:fB}
f_B = \bar f_B(\Omega) - \frac{\lambda c_d}{4\Omega} \int_{-\infty}^{+\infty} d\omega  (K_B[f_0] - K_B[\bar f]),
\end{equation}
where $K_B[f_0] = (1 - 2 f_0(x, \omega + \Omega)) (1 - 2 f_0(x, \omega))$, and $\lambda=1$ for the action above. Below we will also discuss the case of bosons remaining in equilibrium \cite{Nikolaenko2023} (e.g. due to interactions with other degrees of freedom) by setting $\lambda=0$.

\begin{figure}
    \centering
    \includegraphics[width=1.0\columnwidth]{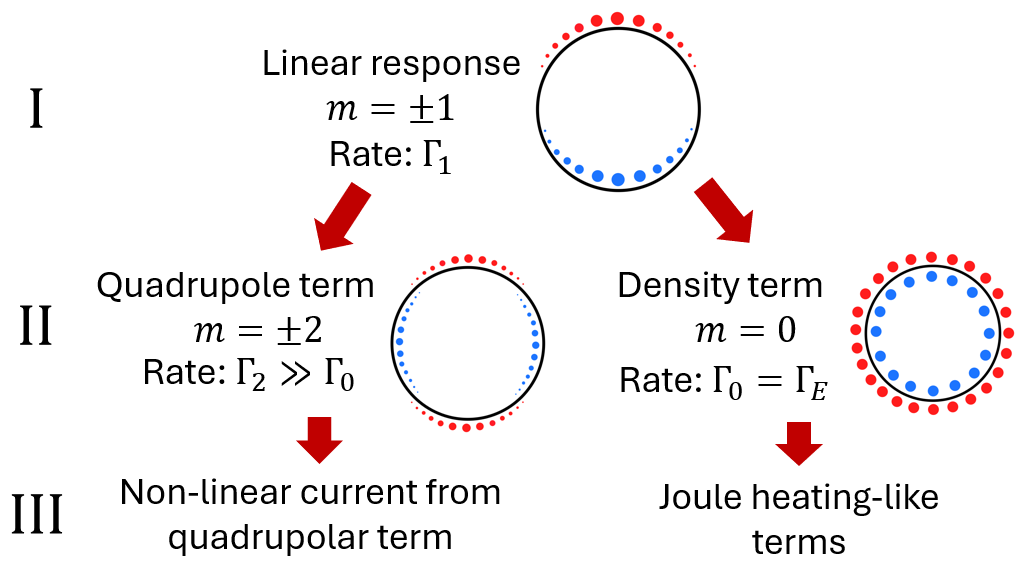}
    \caption{Structure of perturbation theory for Eqs. (\ref{eq:fm},\ref{eq:f0}). Dots represent the angular harmonics $f_m$ of distribution function $f$ generated by electric field. Color represents the sign of perturbation and size --- its magnitude. First order involves only dipolar $m=\pm 1$ harmonics and leads to a linear contribution to current. Second order involves density $m = 0$ and quadrupolar $m = \pm 2$ deformations that do not contribute to current due to inversion symmetry. Both $m = 0$ and $m= \pm 2$ distortions serve as sources to generate $m=\pm1$ distortions in third order, leading to a non-linear contribution to current.
    }
    \label{fig:Perturb}
\end{figure}


\emph{Non-linear responses.} We first analyze nonlinear conductivity perturbatively, solving Eqs. \eqref{eq:fm}, \eqref{eq:f0}, and \eqref{eq:fB} order by order in $\vec E$ (see End Matter and Section III of Supplemental Information for details \cite{Supplement}). 
\sk{In first order, it is sufficient to search for the linear in $E$ contribution to $f$. Electric field $\vec E$ generates  $m = \pm 1$ angular harmonic from the initial $\overline{f}$ in Eq. \eqref{eq:fm}, $f^{1}_m(\omega,T) \propto E v_F \partial_\omega \frac{\overline{f}}{i a(T)\nu +\Gamma_m+g(\omega,T)} \delta_{m,\pm1}$. The current is given by \pv{$\vec j = e \int \frac{d\omega d\theta}{(2\pi)^2 \hbar} \vec k_F (f -\bar f)$ (see Eqs. (\ref{eq:curdef},\ref{eq:fdef}), \cite{Note1})}. Assuming $\Gamma_m\gg g(\omega,T)$, we get the result \cite{Aavishkar2023} for linear conductivity $\sigma(\nu, T) = \frac{e^2}{4 \pi \hbar} \bar W_1(\nu, T)$,} where we defined
\begin{align}\label{eq:W}
    \tilde W_k(\nu, T) = \frac{k_F v_F}{i a(T) \nu+ \Gamma_k + \tilde g(T)},
\end{align}
and $\tilde g(T) \approx \int d\omega \p_\omega \bar f g(\omega, T)$. At the QCP $\tilde g_\mathrm{cr} = \alpha_1 T \tilde \gamma(T)$, where $\tilde \gamma(T) = \gamma(T) + \ln(e/2)$, such that a linear in $T$ corrections to resistivity is obtained. Away from the QCP, the Fermi liquid behavior $\tilde g_\mathrm{FL} = \tilde g_\mathrm{FL}^\prime = \alpha_2 T^2/k_F v_F$ is recovered, where $\alpha_2$ has been defined above.

\sk{In second order in electric field, one needs to compute the deviation from equilibrium for both $f$ and $f_B$ and include non-linearity of collision integrals \cite{Supplement}, which produce quantitative corrections. Solving Eqs. \eqref{eq:fm},\eqref{eq:f0} in second order produces corrections to $m=\pm 2$ and $m=0$ harmonics of $f$.} Involvement of \eqref{eq:f0} warrants  a separate discussion. The collision integral in \eqref{eq:f0} can be shown to possess an eigenstate \cite{Supplement} that depends on $\lambda$ in Eq. \eqref{eq:fB}. Generally, the result takes the form $f^{(2)} \propto W_0(\nu_{ij}, T) \frac{\p}{\p\omega} \left[ W_1(\nu_i, \omega, T) \frac{\p \bar f(\omega)}{\p \omega} \right] $  where $W_0(\nu, T) = \frac{k_F v_F}{i \nu a(T) + g_0(T, \lambda)}$ is independent of $\omega$ and related to the eigenvalue. For the actual model $\lambda=1$, $g_0(T, \lambda=1)=0$, reflecting energy conservation. In this case, the $0$-eigenvalue eigenfunction 
$\delta f_0 \sim \p_T \bar f$, $\delta f_0 \sim \p_T \bar f_B$, which corresponds to response to change of effective temperature (Joule heating). For bosons forced to be in thermal equilibrium $g_0(T, \lambda = 0) = \frac{g^{\prime 2}k_F}{4 \pi v_F} T \sim g(\omega \sim T, T)$, much stronger than what's expected for electron-phonon interactions \cite{allen1987}.

Going further, the structure of kinetic equations implies that in each next order, harmonics with $\delta m  = \pm 1$ are generated (see Fig.~\ref{fig:Perturb}). Most importantly, the $n$-th order correction to distribution function contains $n$ derivatives with respect to  $\omega$, which can act on the nonanalytic (in the $T\to0$ limit) $g(\omega)$, such that $f^{(n)}$ contains a part $\propto \partial^{n-1} g(\omega)$. We show below explicitly that this leads to singular behavior of observable non-linear responses.

\sk{No contribution to current is produced by the $f_{m=\pm2,0}$ reflecting  inversion symmetry, but they both serve as sources for $m=\pm 1$ harmonics at order $E^3$ (Fig. \ref{fig:Perturb}). The third-order current response can correspondingly be decomposed into two contributions $\vec j^{(3)}=\vec j^{(3)}_{0}+\vec j^{(3)}_{\pm 2} $}:
\begin{multline}
    \vec j^{(3)}_{0,\pm 2} = - \frac{e^2}{4 \pi \hbar} A_Q(T) \sum_{ijl} \vec E_l \frac{(\vec E_j \cdot \vec E_i)}{E_0^2} \times 
    \\
    \times \tilde W_{0,\pm 2}(\nu_{ij}) \tilde W^\prime(\nu_i)  \tilde W^{\prime 2}(\nu_{ijl}).
    \label{eq:j3}
\end{multline}
The main difference between the strange metal and the Fermi liquid is the overall magnitude of the response determined by $A_Q(T)$. For strange metal, $A_{Q, \mathrm{cr}} = 2 \alpha_1 k_F v_F/ 3 T \sim T_F/T$, while in a Fermi liquid one obtains $A_{Q, \mathrm{FL}} = 3 \alpha_2/2 \pi^2 \sim 1$. At $T\ll T_F$ strange metal thus has a much stronger (by $T_F/T$) nonlinear response (Fig. \ref{fig:Fan}), formally diverging as $T \to 0$.

The divergence of the third-order response in non-Fermi liquid suggests the breakdown of perturbation theory at low $T$. To understand this regime, we consider higher order terms in the perturbation theory to the set of kinetic equations in question. Examination of the structure of the perturbation series one can see that at the leading order in the limit of small temperature $\vec j^{(n)}  \sim  \vec j^{(n-2)} \vec E^2 / T^2$, where factor of $T$ arises from additional derivatives of $g_\mathrm{cr}(\omega/T)$ (and thus $\Sigma_R(\omega/T)$) over $\omega$. This suggests a universal scaling of the temperature-dependent part of the total current (including all non-linear currents) of the form
\begin{equation}
    \vec j_{\mathrm{tot}}(T) - \vec j_{\mathrm{tot}}(0) = T^2 \vec F\left(\frac{\vec E}{T}\right) \,.
\label{eq:scaling}
\end{equation} This form has similarities to that found near quantum critical points of bosons \cite{Green95,Green13}. \pv{To explore the consequences of strong nonlinearity in the $T\to 0$ limit further, we have also analyzed a simplified version of the kinetic equations with a fixed scattering rate $\Gamma +\alpha_1 T\log\cosh\frac{\omega}{2T} $ for all $m$ while neglecting backaction from bosons. This model can be solved analytically (see End Matter and Supplementary Information, Sec. IV). At low temperatures, $\alpha_1 T \ll \Gamma$, the scaling regime Eq. \eqref{eq:scaling} extends for $\hbar e v_F E \lesssim \Gamma^2/\alpha_1$. For $T\to 0$ (or $\hbar e v_F E \gg \Gamma T$), the nonlinear correction to the current has a nonanalytic form $\propto E^2 \sgn E$, reflecting the non-Fermi liquid scattering $\propto |\omega|$. Finally, for $\hbar e v_F E \gg \Gamma^2/\alpha_1$, the disorder-induced scattering can be neglected and the current behaves as $\propto\sqrt{E}$. For a $\Gamma \sim 1$ meV, $\alpha_1\sim 1$ and $v_F = 3 \cdot 10^5$ m/s \cite{Vishik2010,Hwang2021eo}, the crossover from the scaling regime occurs at $E\sim 50$ V/cm, achievable in modern experiments \cite{katsumi2023}.}


\emph{Discussion and Conclusions:} The results presented above for the 2d-YSYK model can be 
generalized to a broad class of models. The form of the kinetic equation \eqref{eq:fm} relies only on momentum-independence of fermionic self-energy and absence of vertex corrections \cite{Nave2007,arovas2018}. Both are true in single-site dynamical mean field theory (DMFT), which, for the Hubbard model, reproduces strange metal behavior at intermediate temperatures \cite{brown2019bad,cha2020}; moreover, the vertex corrections have been found affect the answer, but only quantitatively \cite{vucicic2019}. Indeed, an enhancement of nonlinear response in a Kondo system has been found in a recent DMFT study \cite{peters2021}. Isotropic strange metal scattering rate has been also reported experimentally \cite{grissonnanche2021linear} and is the cornerstone of the marginal Fermi liquid phenomenology \cite{Varma1989}. While the electron-boson coupling equations (\ref{eq:deltag_main},\ref{eq:I_f0},\ref{eq:fB}) are specific for the YSYK model, their inclusion, does not alter qualitatively the third-order response \eqref{eq:j3}.
\\
The fundamental mechanism for the divergent nonlinear response we identify is the non-analytic behavior of the self-energy ${\rm Im} \Sigma_R''(\omega,T\to 0)\propto \frac{1}{T}$. Thus, our conclusions can be expected to hold for a broad class of models where self-energy is local and non-analytic, such as for $q=0$ quantum critical points \cite{Sachdev_2011,lohnheysen2007}, e.g. nematic \cite{shibauchi2014quantum}. Specifically, for ${\rm Im} \Sigma_R(\omega)_{T\to0} \propto \omega^\alpha$, our results suggest a nonlinear response $\sigma^{(3)} \propto T^{\alpha-2}$ at finite $T$.

Even more generally, at low frequencies the $E_F/T$ enhancement can be understood for the nonlinear response arising from Joule heating ($m=0$ harmonic taken at second order). The temperature change due to heating is $\Delta T \propto  V^2/(R C_{el})$, which leads to the nonlinear correction $R(T)\approx R(T_0)+R'(T) \Delta T$, such that $\Delta T\propto V^2$. This is the basis of the so-called 3-$\omega$ method \cite{cahill1987}. The difference between strange metal and Fermi liquid is reduced then to $R'(T)$ being constant (strange metal) or of the order $T/T_F$ (Fermi liquid). The same argument applied to a non-Fermi liquid metal with $R\propto T^{\alpha}$, would result in an enhancement $(T_F/T)^{\alpha-1}$ for the strange metal. The argument above points to the intimate relation between inelastic scattering, temperature dependence of resistivity and nonlinear responses (see Fig. \ref{fig:Fan}). Since the conventional definition of strange metal is based on the second property, our work opens another perspective on studying this phenomenon. \sk{Using this argument, one can connect electronic single sheet heat capacity $C_{el}^{(1)}$ from heating to microscopic quantities that appear in Eq. \eqref{eq:j3} through $C_{el}^{(1)}(T) \sim e^2 E_0^2 T/8 \pi E_F^3$.}


Our results also highlight that nonlinear response contains much more information about the correlated electrons, than the linear one. In present model, scattering rates for all angular harmonics $m$ are equivalent; however, this no longer holds when lattice symmetry is taken into account \cite{Supplement}. Instead, different scattering rates should be attributed to different irreducible representations of the point group. For example, for $D_{4h}$ group appropriate for square lattice materials, such as cuprates, instead of $m=\pm 2$ Fermi surface deformations, there would be $B_{1g}$ and $B_{2g}$ ones, characterized by different scattering rates. Those scattering rates can be extracted from the frequency dependence of the third-order response, Eq. \eqref{eq:j3}. The main obstacle on the way to successfully study these quadrupolar relaxation rates is that $\vec j^{(3)}_{0} $ is generally larger then $\vec j^{(3)}_{\pm 2} $, since the relaxation rate $g_0$ of the energy density is much smaller than the relaxation of quadrupolar harmonics $\Gamma_2$. However, exploiting polarization dependence of the response can allow to isolate the quadrupolar part. Assuming  $x$- and $y$- directions are chosen to be along the principal axes, $\vec E$ applied along the axes would cause only $B_{1g}$-type quadrupolar response. When $\vec E$ applied along diagonals, only $B_{2g}$-type quadrupolar response will be triggered. In both cases $\vec j^{(3)}_0, \vec j^{(3)}_{\pm 2} \parallel \vec E$, but only $\vec j^{(3)}_0$ would be the same, such that subtracting two results allows to isolate the quadrupolar-mediated part. Furthermore, when $E$ is not directed along the main axes or diagonal, the component of $\vec j^{(3)}_{\pm 2}$ perpendicular to $\vec E$ is solely quadrupolar-mediated (see Appendix V in Supplemental Information \cite{Supplement} for full expressions).

We now discuss the application of our results to known strange metals. 
\sk{For cuprates, we estimate the non-linear conductivity associated with $\vec j^{(3)}_{\pm 2}$ (non heating) for LSCO
assuming all $\Gamma_{m \neq 0}$ are equal with the set of parameters taken at $T\sim40$ K from experiments \cite{doi:10.1126/science.1165015,Legros2019,semenenko_magnetization_2020}
(see details in Sec. VI of Supplemental Information \cite{Supplement}).
We obtain $\sigma^{(3)}\sim 1 \cdot 10^{-8} A_Q$~($\Omega$~m)$^{-1}$/(V/m)$^{-2}$, or $\sigma^{(3)}/\sigma^{(1)} \sim 5 \cdot 10^{-15} A_Q$~(V/m)$^{-2}$. }
For a Fermi liquid $A_Q\sim 1$ with the same parameters, non-linear response becomes comparable to linear at $\sim 150 \, \mathrm{kV/cm}$, for a non-Fermi liquid with $A_Q \sim T_F/T \sim 50$ only fields of strength $20 \, \mathrm{kV/cm}$ are required.
In modern THz experiments, \sk{recently conducted on a number of correlated metals \cite{katsumi2023,barbalas2023,chaudhuri2025},} field strengths well in excess of that can be generated \cite{chaudhuri2025}.

\sk{For experiments with pulsed field of duration $\delta t \ll \hbar/\Gamma_0$, heating-related nonlinear response can be estimated from the temperature increase due to Joule heating \cite{Supplement}. For a pulse of $\sim 1$ ps \cite{barbalas2023,chaudhuri2025} in LSCO with the same parameters as above, we get $\sigma^{(3)}_J \sim 6 \cdot 10^{-6}$~($\Omega$~m)$^{-1}$/(V/m)$^{-2}$, or $\sigma^{(3)}_J/\sigma^{(1)} \sim 3 \cdot 10^{-12} \, (\mathrm{V}/\mathrm{m})^{-2}$, which can be appreciable for fields $\sim 6 \,\mathrm{kV}/\mathrm{cm}$ \cite{katsumi2023}. In the Fermi liquid regime, the estimate is noticeably smaller, chiefly due to decrease of $d \rho/dT$ by a factor of $5$.}

In conclusion, we demonstrated that a strong $T_F/T$ enhancement of nonlinear responses is a characteristic feature of strange metals with respect to Fermi liquids. Our results suggest the existence of an $E/T$ scaling behavior of nonlinear conductivity of strange metals at low temperatures. Third order responses in particular have been shown to contain information about relaxation time of quadrupolar distortions of the Fermi surface that can be deduced in experiments with controlled field orientation. Our estimates \sk{show that the predicted phenomena are well within reach of modern THz experiments, and are in qualitative agreement with features observed recently in \cite{chaudhuri2025}.}

\emph{Acknowledgements}
We acknowledge useful discussions with D. Natelson and E. Konig. We also acknowledge useful discussions with Haoyu Guo, in particular, pointing out the importance of non-linear terms in the kinetic equation \cite{Haoyu24}. This research was supported by the U.S. National Science Foundation grant No. DMR-2245246 and by the Simons Collaboration on Ultra-Quantum Matter which is a grant from the Simons Foundation (651440, S.S.)


\bibliography{NonLin}
\onecolumngrid
 \newpage
\begin{center}
{\bf\large End Matter}
\end{center}

\appendix*
 \setcounter{equation}{0}

 \section{Keldysh formalism}

 Here we present an overview of the kinetic equation derivation for the YSYK model; details of calculations with appropriate references can be found in Supplemental Information \cite{Supplement}. We first obtain an effective action for field on the Keldysh time contour \cite{Kamenev_2011}. To achieve that we first rewrite the partition function $Z = \int D[\psi_{\pm}, \psi^\dagger_{\pm}, \phi_{\pm}] \; e^{i S_\mathrm{tot}}$,
where $S_\mathrm{tot}$ is given by Eq. 7 in the main text. Specifically, we use Hubbard-Stratonovich transformation introducing introduce bi-local effective fields \cite{Guo2021,Guo2022,Aavishkar2023} $i G_{\alpha \beta}(x, x^\prime) = \frac{1}{N} \sum_{i = 1}^N \psi_{\alpha i}(x) \psi^\dagger_{\beta i}(x^\prime)$ and $i D_{\rho \lambda} (x, x^\prime) = \frac{1}{N} \sum_{i = 1}^N \phi_{\rho i}(x) \phi_{\lambda i}(x^\prime)$,
where $x = (t, \vec r)$, latin indices correspond to SYK field flavors and greek indices correspond to Keldysh contour temporal components. Additionally we introduce fields $\Sigma_{\alpha \beta}(x, x^\prime)$ and $\Pi_{\rho \lambda}(x, x^\prime)$ that serve as Lagrange multipliers to $G_{\alpha\beta}$ and $D_{\rho\lambda}$ correspondingly. After performing that procedure, the resulting partition function is averaged over quenched disorder (Eq. 5,6) via $    \langle Z \rangle = \int D[v, g^\prime] Z(v, g^\prime) 
    e^{- 2 \sum_{i\leq j}\frac{|v_{ij}|^2(\vec r)}{v^2}}
    e^{- 2 \sum_{i\leq j} \frac{2 |g^{\prime}_{ijl}|^2(\vec r)}{g^{\prime 2}}}.$ 
Integrating over the original fermion/boson fields and disorder  results in effective action (note that we work in $2+1$ dimensions so $\int d^3 x$ is over both space and time):
\begin{multline} \label{eq:eff_action}
    \frac{S_\mathrm{eff}}{N} = - i \Tr \ln \left( G_0^{-1} - \Sigma \right)
    + \frac{i}{2} \Tr \ln \left( D_0^{-1} - \Pi\right)
    + 
    i \int d^3x \, d^3 x^\prime \left( \frac{1}{2} \Pi_{\lambda \rho}(x^\prime, x) D_{\rho \lambda}(x, x^\prime) - \Sigma_{\alpha \beta}(x^\prime x) G_{\beta \alpha}(x, x^\prime) \right)
    \\
    + \frac{i v^2}{2} \int d^3 x \, d^3 x^\prime \; \delta(\vec r - \vec r^\prime) \tilde \delta_{F,\alpha \beta} \tilde \delta_{F,\mu \nu} G_{\nu \alpha}(x^\prime, x) G_{\beta \mu}(x, x^\prime)
    - 
    \frac{c^2g^{\prime 2}}{2} \int d^3x \, d^3 x^\prime\; \tilde \delta_{\alpha\beta\rho} \tilde \delta_{\mu \nu \lambda} G_{\nu \alpha} (x^\prime, x) G_{\beta \mu}(x, x^\prime) D_{\rho \lambda}(x, x^\prime). 
\end{multline} The coefficients $\tilde \delta_{F, \alpha \beta}$ and $\tilde \delta_{\alpha \beta \rho}$ are $
    \tilde \delta_{F, \alpha \beta} =
    \begin{bmatrix}
        1 & 0 \\
        0 & -1
    \end{bmatrix},
    \quad
    \tilde \delta_{\alpha \beta 1}=
    \begin{bmatrix}
        1 & 0 \\
        0 & 0
    \end{bmatrix},
    \quad
    \tilde \delta_{\alpha \beta 2}=
    \begin{bmatrix}
        0 & 0 \\
        0 & -1
    \end{bmatrix},
$
and $G_0^{-1}$ and $D_0^{-1}$ are bare inverse fermion $\psi$ and boson $\phi$ Green's functions defined by original action.
Since the whole action is proportional to $N$, we apply a large-$N$ expansion that leads to the equations of motion for fields $G, D, \Sigma$, and $\Pi$ for the action that correspond to the saddle point of the action in Eq. \eqref{eq:eff_action}.
Varying over the self-energies results in the Dyson equations $G_{\alpha \beta}(x, x^\prime) = [ ( G_0^{-1} - \Sigma )^{-1}]_{\alpha \beta}(x, x^\prime)$, and  $D_{\rho \lambda}(x, x^\prime) = [ ( D_0^{-1} - \Pi )^{-1}]_{\rho \lambda} (x, x^\prime)$.
Varying the action over $G$ and $D$ results in 1-loop self-consistent expressions for self-energies $\Sigma$ and $\Pi$:
\begin{align} \label{eq:Sigma}
    i \Sigma_{\alpha \beta} (x, x^\prime) = i v^2 \delta(\vec r - \vec r^\prime) \tilde \delta_{\alpha \mu} \tilde
    \delta_{\nu \beta} G_{\mu \nu}&(x, x^\prime)    
    - 
    \frac{c^2 g^{\prime 2}}{2} \delta(\vec r - \vec r^\prime) \tilde \delta_{\alpha \nu \rho} \tilde \delta_{\mu \beta \lambda} G_{\nu \mu}(x, x^\prime)  
    \left( D_{\rho \lambda}(x, x^\prime) + D_{\lambda \rho}(x^\prime, x) \right),
    \\\label{eq:Pi}
    i \Pi_{\lambda \rho}(x, x^\prime) &= c^2g^{\prime 2} \delta(\vec r - \vec r^\prime) \tilde\delta_{\alpha\beta\rho} \tilde \delta_{\mu \nu \lambda} G_{\nu \alpha}(x, x^\prime) G_{\beta \mu}(x^\prime, x).
\end{align}

These equations, together with Green's functions definitions constitute, in principle, a full set of equations describing the system. \pv{The utility of large N approximation consists in the absence of vertex corrections in these equations, while disorder in the couplings ensures momentum-independence of self energies. Similar equations will therefore hold in other theories or approximations that have these two features.} One can simplify these equaitions bringing them to a familiar form of kinetic equation for distribution functions. We perform a Keldysh rotation \cite{Kamenev_2011} from the original contour variables to 
$
    \begin{bmatrix}
        \phi_1 \\
        \phi_2
    \end{bmatrix}
    =
    \frac{1}{\sqrt{2}}
    \begin{bmatrix}
        1 & 1 \\
        1 & -1
    \end{bmatrix}
    \begin{bmatrix}
        \phi_{+} \\
        \phi_{-}
    \end{bmatrix},
    \quad
            \begin{bmatrix}
        \psi_1 \\
        \psi_2
    \end{bmatrix}
    =
    \frac{1}{\sqrt{2}}
    \begin{bmatrix}
        1 & 1 \\
        1 & -1
    \end{bmatrix}
    \begin{bmatrix}
        \psi_{+} \\
        \psi_{-}
    \end{bmatrix},
    \quad
        \begin{bmatrix}
        \psi_1^\dagger \\
        \psi_2^\dagger
    \end{bmatrix}
    =
    \frac{1}{\sqrt{2}}
    \begin{bmatrix}
        1 & -1 \\
        1 & 1
    \end{bmatrix}
    \begin{bmatrix}
        \psi_{+}^\dagger \\
        \psi_{-}^\dagger
    \end{bmatrix}.
$
Fields with subscript $+$ and $-$ correspond t forward- and backward-propagating parts of the contour correspondingly. We further introduce $G^< = (G_K - G_R + G_A)/2$ and $D^< = (D_K - D_R + D_A)$ instead of $G_K$ and $D_K$, as they allow to define the fermionic $f$ and bosonic $f_B$ quantities that play a role of occupation functions, Eq. 10 and 11 of the main text.  $G^<$ and $D^<$ are understood as Wigner-transformed correlation functions where $\omega$ and $\Omega$ correspond to the excitation energies, while $\vec k$ and $\vec q$ correspond to the excitations' momenta. The Dyson equations for $G^<$ and $D^<$ can be rewritten in real space
\begin{align}\label{eq:GlessKinCoord}
    [G_0^{-1}; G^<] &= \Sigma_R \circ G^< - G^< \circ \Sigma_A
    + \frac{1}{2} \left( \Sigma_K + \Sigma_A - \Sigma_R \right) \circ G_A - \frac{1}{2} G_R \circ \left(  \Sigma_K + \Sigma_A - \Sigma_R \right)
    \\\label{eq:DlessKinCoord}
    [D_0^{-1}; D^<] &= \Pi_R \circ D^< - D^< \circ \Pi_A
    + \frac{1}{2} \left( \Pi_K + \Pi_A - \Pi_R \right) \circ D_A - \frac{1}{2} D_R \circ \left(  \Pi_K + \Pi_A - \Pi_R \right).
\end{align}
To evaluate the self-energies, we perform a Wigner-transform of Eqs. \eqref{eq:Sigma} and \eqref{eq:Pi} and use retarded Green's function obtained previously \cite{Guo2021}: $G_R = (\omega - v_F k - i \Im \Sigma_R(\omega, t))^{-1}$ and $D_R = (- \Omega^2 + m^2 + q^2 - 2 i \Omega c_d)$. The self-energies can be expressed a functionals of $f$ and $f_B$ only, rather than full $G^<$ and $D^<$, by performing the momentum integrals. \pv{This step generalizes to any models without vertex corrections and where self-energies are momentum independent; for YSYK the result is:}
\begin{align}
    \Pi_R(x, \omega) &= -i \frac{g^{\prime 2} k_F^2}{v_F^2} \int_{-\infty}^{+\infty} \frac{d \omega}{2 \pi} \int_0^{2 \pi} \frac{d \theta}{2 \pi} \left[ f(x, \omega, \theta) - f(x, \omega + \Omega, \theta) \right]
    \\
    \Pi_K(x, \omega) &= i\frac{g^{\prime 2} k_F^2}{2 v_F^2} \int_{-\infty}^{+\infty} \frac{d \omega}{2 \pi} \int_0^{2\pi} \frac{d\theta}{2 \pi} \frac{d \theta^\prime}{2 \pi} \left[ (1 - 2 f(x, \omega + \Omega, \theta))(1 - 2 f(x, \omega, \theta^\prime)) - 1 \right].
    \\
    \Im \Sigma_R(x, \omega) &= - \frac{\Gamma}{2} - \frac{g^{\prime 2} k_F}{v_F} \int \frac{d \Omega}{2 \pi} \int \frac{d \theta}{2 \pi} B(\Omega) \left[ f_B(x, \Omega, \theta) - f(x, \omega - \Omega, \theta) + 1 \right]
    \\\label{eq:ReSigmaR}
    \Re \Sigma_R(x, \omega) &= \frac{g^{\prime 2}k_F}{2 v_F} \int \frac{d \Omega}{2 \pi} \int \frac{d \theta}{2 \pi} B^\prime(\Omega) (2 f(x, \omega - \Omega, \theta) - 1),
    \\
        \Sigma_K(x, \omega) &= 4 i \Gamma\int \frac{d\theta}{2 \pi} f(x, \omega, \theta) - 2 i \Gamma+ i \frac{g^{\prime2} k_F}{v_F} \int\frac{d\Omega}{2 \pi} \int \frac{d\theta d\theta^\prime}{(2 \pi)^2} B(\Omega) (2 f_B(\Omega, \theta^\prime)+ 1) (f(\omega+ \Omega, \theta) + f(\omega - \Omega, \theta) -1).
\end{align}
Performing the Wigner transformation on Eqs. \eqref{eq:GlessKinCoord} and \eqref{eq:DlessKinCoord}, substituting the self-energies above yields the system of Boltzmann-like kinetic equations for fermions (Eq. 12-15) and bosons (Eq. 16, where we additionally assumed strong damping $c_d \gg \frac{\partial_t f_B}{f_B-\overline{f}_B} $).

 \section{Non-perturbative solution of a simplified model}

 In the main text we have shown that the kinetic equations  (\ref{eq:fm},\ref{eq:f0},\ref{eq:fB}) for YSYK model result in a divergent nonlinear response in the $T\to 0$ limit, indicating breakdown of the perturbation theory. The divergence originated from the nonanalyticity of the scattering rate $g(\omega,T)$ in \eqref{eq:fm} the $T\to 0$ limit. Based this observation, we consider of simplification of Eq. (\ref{eq:fm},\ref{eq:f0},\ref{eq:fB}), that also has a $1/T$ divergence, but allows for a full analytical solution, non-perturbative in $E$. We assume the scattering rate to be the same for all $m$ and neglect the backaction of bosons, Eq. \eqref{eq:fB}. The problem then reduces to a single kinetic equation:
\begin{equation}
\begin{gathered}
    e v_F E \cos \theta \, \p_\omega f + \Gamma(\omega) f = \Gamma(\omega) \bar f,
    \end{gathered}
\end{equation}
where $    \Gamma(\omega) = \Gamma_0+\alpha_1 T \log \cosh[\omega/2 T]$ and $\bar f(\omega,T) = (e^{\omega/T}+1)^{-1}$. The current for this model can be found analytically. It is directed along ${\bf E}$ with its magnitude $j$ given by:
\begin{equation}\label{em:eq:j}
\begin{gathered}
    j = \frac{ e k_F}{8\pi^2} \int_{-\pi/2}^{\pi/2} d\theta \cos \theta \int_{-\infty}^\infty d\omega 
    \int_{0}^{\infty} d\delta 
 \exp \left[ - \frac{\Gamma_0 \delta + F(\omega,\delta)}{e v_F E \cos \theta} \right]
\frac{1}{ T\cosh^2\frac{\omega}{2T}},
\end{gathered}
\end{equation}
where $F(\omega,\delta) = \int_{\omega}^{\omega+\delta} d\omega'\alpha_1 T \log \cosh[\omega'/2 T]$. Here we will focus on the low-temperature $\alpha_1 T\ll \Gamma_0$ limit; details of the calculations, results for the high-temperature limit  $\Gamma_0\ll \alpha_1 T$ and additional numerical results are presented in Supplementary Material, Sec. IV. 

At sufficiently low fields, $\delta \sim \frac{e v_F E \cos \theta}{\Gamma_0} \ll T \sim \omega$ and one nonlinear response can be obtained perturbatively by a Taylor expansion of $F[\omega,\delta]$ in $\delta$. Lowest-order nonlinear response arises from the cubic term in the expansion $F^{(3)}(\omega,\delta) = \frac{\alpha_1 \delta^3}{24 T \cosh^2[\omega/2 T]}$ and leads to $j^{(3)} \propto 1/T$, as in the full model studied in the main text. Furthermore, one notices that $F(\omega, \delta) = T^2 \phi\left( \frac{\delta}{T},\frac{\omega}{2T}\right)$. As long as $F(\omega, \delta) \ll e E v_F \cos \theta$, we can use $e^{-F}\approx 1-F$. This results in the nonlinear correction to current taking on a scaling form, $j_{nonlin} \propto \frac{T^2}{\Gamma_0} f_{sc} \left(\frac{E}{\Gamma_0 T}\right)$, in agreement with Eq. \eqref{eq:scaling}. At finite temperatures, we found that an empirical scaling form $j = \sigma(T) T^2  f(E \sigma(T) /T)$, where $\sigma(T)$ is the linear conductivity, holds for an extended parameter range (see SM).

Eq. \eqref{em:eq:j} can be analyzed also directly at $T=0$, where the integral over $\omega$ amounts to a delta function and $F(\omega=0,
\delta)= \alpha_1 \delta^2/2$. For $ev_F E \ll \Gamma_0^2/\alpha_1$ perturbative evaluation yields an $\propto E^2$ correction to $j$. Therefore, at $T\to 0$ the $1/T$ singularity in $j^{(3)}$ is avoided, because the lowest-order nonlinear correction becomes non-analytic: $j^{(nonlin)}_{T\to 0, E\to 0} \propto E^2$, but directed along ${\bf E}$. This implies, e.g., for ${\bf E}\parallel x$, $j^{(nonlin)}_{T\to 0, E\to 0} \propto E_x^2 \sgn E_x$. This result corresponds to the limit $f_{sc}(x\to \infty)$ and therefore we get $f_{sc}(x\to \infty) \propto x^2$. We notice that this can be contrasted with the behavior of a Fermi liquid, that can be simulated with $\Gamma(\omega) = \Gamma_0+\alpha_{FL} \omega^2$, leading to an analytic $j^{(nonlin,FL)}_{T\to 0, E\to 0} \propto E_x^3$.

A final question pertains regarding the extent of the $\frac{E}{\Gamma_0 T}$ scaling regime in temperature and electric field. At low temperatures $\alpha_1 T \ll \Gamma_0$ and at low $E$ the characteristic $\delta \sim e E v_F \cos \theta/\Gamma_0 \ll T$. Then, $F(\omega,\delta) \sim \alpha_1 T \delta \ll e E v_F \cos \theta$. This condition will thus break only when typical $\delta$ becomes of the order $T$ or larger. Assuming the latter, we can estimate $F(\omega,\delta \gg T) \approx \alpha \delta^2 /2$. The scaling will break down when $F(\omega,\delta \gg T)$, so that $\alpha_1 \delta^2 \sim e E v_F $. Using  $\delta \lesssim e E v_F /\Gamma_0$, we find that this will occur at $eEv_F \sim \Gamma_0^2/\alpha_1$. Therefore at $ T \ll \Gamma_0/\alpha_1$, the scaling regime extends to field  $E \lesssim \Gamma_0^2/(e v_F\alpha_1)$. In the Supplementary Information we show that for large temperatures $T\gg \Gamma_0/\alpha_1$, the scaling regime also exists and extends to $E \lesssim \sqrt{\Gamma_0^3 T/\alpha_1}/(e v_F)$.

\newpage


\section*{Supplementary material}

# Supplemental Information for
# 'Strong non-linear response of strange metals'

Serhii Kryhin,[1, *] Subir Sachdev,[1, †] and Pavel A. Volkov[2, ‡]

[1]*Department of Physics, Harvard University, Cambridge MA 02138, USA*
[2]*Deparment of Physics, University of Connecticut, Storrs, CT 06269, USA*
(Dated: February 25, 2025)

## I. LARGE-$N$ EXPANSION OF THE YUKAWA-SYK MODEL IN THE KELDYSH FIELD THEORY

To derive the self-consistent large-$N$ expansion in the Yukawa-SYK model of interest, we adopt a routine procedure previously described in [1–3] for Matsubara field theory. The only difference for Keldysh field theory, that we use here, is the additional index for the quantum fields, indicating the forward and reverse part of the closed time contour; in this we follow [4]. We average over the ensemble of gaussian-distributed random disorder $v_{ij}(\mathbf{r})$ and random coupling $g_{ijl}(\mathbf{r})$ with the averages and variances given by

$$\langle v_{ij}(\mathbf{r}) \rangle = 0, \tag{1}$$

$$\langle g_{ijl}(\mathbf{r}) \rangle = 0, \tag{2}$$

$$\langle v_{ij}(\mathbf{r}) \rangle = \langle g'_{ijl}(\mathbf{r}) \rangle = 0, \tag{3}$$

$$\langle v^*_{ij}(\mathbf{r}) v_{ab}(\mathbf{r}') \rangle = v^2 \delta_{ia} \delta_{jb} \, \delta(\mathbf{r} - \mathbf{r}'), \tag{4}$$

$$\langle g^{*\prime}_{ijl}(\mathbf{r}) g^{*\prime}_{abc}(\mathbf{r}') \rangle = g'^2 \delta_{ia} \delta_{jb} \delta_{lc} \, \delta(\mathbf{r} - \mathbf{r}'), \tag{5}$$

where the averaging for any observable $f$ is done by

$$\langle f \rangle = \int D[v, g'] f(v, g') e^{-2 \sum_{i \leq j} \frac{|v_{ij}|^2(\mathbf{r})}{v^2}} e^{-2 \sum_{i \leq j} \frac{2|g'_{ijl}|^2(\mathbf{r})}{g'^2}}. \tag{6}$$

Following the usual procedure of $\Sigma - G$ method [1–3], we average over the partition function $Z$ of the original theory given by

$$Z = \int D[\psi_\pm, \psi^\dagger_\pm, \phi_\pm] \, e^{iS_{\text{tot}}}, \tag{7}$$

where $S_{\text{tot}}$ is given by Eq. 8 in the main text. Averaging over the random variables $v_{ij}, g'_{ijl}$ for $Z$ results in

$$\langle Z \rangle = \int D[\psi_\pm, \psi^\dagger_\pm, \phi_\pm] \, e^{iS_\phi + iS_\psi} \times$$

$$\times \exp\left[ -\frac{v^2}{2N} \sum_{ij} \int dx \, dx' \, \delta(\mathbf{r} - \mathbf{r}') \psi^\dagger_{\alpha i} \psi_{\beta j}(x) \psi^\dagger_{\mu j} \psi_{\nu i}(x') \delta_{F, \alpha\beta} \delta_{F, \mu\nu} \right] \times$$

$$\times \exp\left[ -\frac{g'^2 c^2}{2N^2} \sum_{ijl} \int dx \, dx' \, \delta(\mathbf{r} - \mathbf{r}') \psi^\dagger_{\alpha i} \psi_{\beta j} \phi_{\lambda l}(x) \psi^\dagger_{\mu j} \psi_{\nu i} \phi_{\rho l}(x') \tilde{\delta}_{\alpha\beta\lambda} \tilde{\delta}_{\mu\nu\rho} \right], \tag{8}$$

where we introduced the new notation as following. We denote the $\pm$ time contour index as a Greek index: field $\psi_{i+} = \psi_{i1}$, and $\psi_{i-} = \psi_{i2}$. The coefficients $\tilde{\delta}_{F,\alpha\beta}$ and $\tilde{\delta}_{\alpha\beta\rho}$ are

$$\tilde{\delta}_{F,\alpha\beta} = \begin{bmatrix} 1 & 0 \\ 0 & -1 \end{bmatrix}_{\alpha\beta} \qquad \tilde{\delta}_{\alpha\beta 1} = \begin{bmatrix} 1 & 0 \\ 0 & 0 \end{bmatrix}_{\alpha\beta} \qquad \tilde{\delta}_{\alpha\beta 2} = \begin{bmatrix} 0 & 0 \\ 0 & -1 \end{bmatrix}_{\alpha\beta} \tag{9}$$

---
* skryhin@g.harvard.edu
† sachdev@g.harvard.edu
‡ pavel.volkov@uconn.edu

To reproduce Eq. 8 of the main text.

We proceed by defining bilocal in spacetime fields $G_{\alpha\beta}(x,x')$ and $D_{\rho\lambda}(x,x')$ as

$$iG_{\alpha\beta}(x,x') = \frac{1}{N}\sum_{i=1}^{N}\psi_{\alpha i}(x)\psi^{\dagger}_{\beta i}(x'),\tag{10}$$

$$iD_{\rho\lambda}(x,x') = \frac{1}{N}\sum_{i=1}^{N}\phi_{\rho i}(x)\phi_{\lambda i}(x'),\tag{11}$$

where $x = (t,\mathbf{r})$. This definition is consistent with the definition of Green's functions in real time up to averages. We can now decouple the interaction terms in the action, Eq.(8) by introducing Lagrange multipliers $\Sigma_{\alpha\beta}(x,x')$ and $\Pi_{\alpha\beta}(x,x')$ via $\int D[\Sigma]\exp\left[-i\Sigma\left(iG - \frac{1}{N}\sum_{i=1}^{N}\psi_{\alpha i}(x)\psi^{\dagger}_{\beta i}(x')\right)\right] = \delta\left[-i\Sigma\left(iG - \frac{1}{N}\sum_{i=1}^{N}\psi_{\alpha i}(x)\psi^{\dagger}_{\beta i}(x')\right)\right]$. The transformation leads to an action quadratic in $\psi$ and $\phi$. Integrating out the original fields results in effective action (note that we work in $2+1$ dimensions so $\int d^3x$ is over both space and time):

$$
\begin{aligned}
\frac{S_{\text{eff}}}{N} = {}& -i\,\text{Tr}\,\ln\left(G_0^{-1} - \Sigma\right) + \frac{i}{2}\,\text{Tr}\,\ln\left(D_0^{-1} - \Pi\right) \\
&+ i\int d^3x\,d^3x'\left(\frac{1}{2}\Pi_{\lambda\rho}(x',x)D_{\rho\lambda}(x,x') - \Sigma_{\alpha\beta}(x'x)G_{\beta\alpha}(x,x')\right) \\
&+ \frac{iv^2}{2}\int d^3x\,d^3x'\,\delta(\mathbf{r}-\mathbf{r}')\tilde{\delta}_{F,\alpha\beta}\tilde{\delta}_{F,\mu\nu}G_{\nu\alpha}(x',x)G_{\beta\mu}(x,x') \\
&- \frac{c^2g'^2}{2}\int d^3x\,d^3x'\,\tilde{\delta}_{\alpha\beta\rho}\tilde{\delta}_{\mu\nu\lambda}G_{\nu\alpha}(x',x)G_{\beta\mu}(x,x')D_{\rho\lambda}(x,x').
\end{aligned}\tag{12}
$$

This expression consists of several parts. The traces in the first line of Eq. (12) come from integrating out the fields $\psi$ and $\phi$. The expressions for $G_{0,\alpha\beta}$ and $D_{0,\rho\lambda}$ are the corresponding bare fermion and boson Green's functions of our model. The terms with "self-energies" $\Sigma$ and $\Pi$ arise from the Lagrange multiplier term. The term in the third line corresponds to the potential disorder, and finally, the last line corresponds to the interaction averaged over disorder realizations.

Since the whole action is proportional to $N$, we apply a large-$N$ expansion that leads to the equations of motion for fields $G, D, \Sigma$, and $\Pi$ for the action that correspond to the saddle point of the action in Eq. (12).

Varying over the self-energies results in the Dyson equations

$$G_{\alpha\beta}(x,x') = \left[\left(G_0^{-1} - \Sigma\right)^{-1}\right]_{\alpha\beta}(x,x'),\tag{13}$$

$$D_{\rho\lambda}(x,x') = \left[\left(D_0^{-1} - \Pi\right)^{-1}\right]_{\rho\lambda}(x,x').\tag{14}$$

Varying the action over $G$ results in

$$i\Sigma_{\alpha\beta}(x,x') = i\Sigma_{v,\alpha\beta}(x,x') + i\Sigma_{g',\alpha\beta}(x,x'),\tag{15}$$

The corresponding $v$ and $g'$ contributions are

$$i\Sigma_{v,\alpha\beta}(x,x') = iv^2\delta(\mathbf{r}-\mathbf{r}')\tilde{\delta}_{\alpha\mu}\tilde{\delta}_{\nu\beta}G_{\mu\nu}(x,x')\tag{16}$$

$$i\Sigma_{g',\alpha\beta}(x,x') = -\frac{c^2g'^2}{2}\delta(\mathbf{r}-\mathbf{r}')\tilde{\delta}_{\alpha\nu\rho}\tilde{\delta}_{\mu\beta\lambda}G_{\nu\mu}(x,x')\left(D_{\rho\lambda}(x,x') + D_{\lambda\rho}(x',x)\right).\tag{17}$$

Varying over $D$ results in

$$i\Pi_{\lambda\rho}(x,x') = c^2g'^2\delta(\mathbf{r}-\mathbf{r}')\tilde{\delta}_{\alpha\beta\rho}\tilde{\delta}_{\mu\nu\lambda}G_{\nu\alpha}(x,x')G_{\beta\mu}(x',x).\tag{18}$$

The Eqs (15) and (18) correspond to 1-loop self-consistent large-$N$ expansion in the Keldysh theory. These results are used to construct a system of kinetic equations that would describe $v - g'$ model, but first a Keldysh rotation needs to be performed.

We construct a Keldysh rotation in a manner similar to the method employed by Kamenev in [4]. We define the Keldysh rotation for bosons by a transformation of the form

$$\begin{bmatrix}\phi_1\\\phi_2\end{bmatrix} = \frac{1}{\sqrt{2}}\begin{bmatrix}1 & 1\\1 & -1\end{bmatrix}\begin{bmatrix}\phi_+\\\phi_-\end{bmatrix} = \Lambda\begin{bmatrix}\phi_+\\\phi_-\end{bmatrix}.\tag{19}$$

The Keldysh rotation for fermions is defined in a more complicated but consistent with literature manner, since hermitian conjugated fields transform differently from their counterparts:

$$\begin{bmatrix} \psi_1 \\ \psi_2 \end{bmatrix} = \frac{1}{\sqrt{2}} \begin{bmatrix} 1 & 1 \\ 1 & -1 \end{bmatrix} \begin{bmatrix} \psi_+ \\ \psi_- \end{bmatrix} = U \begin{bmatrix} \psi_+ \\ \psi_- \end{bmatrix}, \qquad \begin{bmatrix} \psi_1^\dagger \\ \psi_2^\dagger \end{bmatrix} = \frac{1}{\sqrt{2}} \begin{bmatrix} 1 & -1 \\ 1 & 1 \end{bmatrix} \begin{bmatrix} \psi_+^\dagger \\ \psi_-^\dagger \end{bmatrix} = V \begin{bmatrix} \psi_+^\dagger \\ \psi_-^\dagger \end{bmatrix}. \tag{20}$$

The Keldysh rotation defined by matrices $\Lambda$, $U$, and $V$ naturally induces the Keldysh rotation for the Green's functions $G_{\alpha\beta}$ and $D_{\rho\lambda}$ defined by Eqs. (10) and (11). Since the Keldysh rotation is nearly a basis change for the fields, the action $S_{\text{eff}}$ has to be invariant under the rotation. In turn, this naturally induces the transformation for $\Sigma_{\alpha\beta}$ and $\Pi_{\rho\lambda}$.

If the theory is self-consistent, the four components of the Green's function matrix are linearly dependent. Keldysh rotation is designed to explicitly eliminate one of the components by a change of basis. With our choice of the rotation in Eqs. (19) and (20) we expect the structure of the Green's functions after Keldysh rotation to be

$$G_{\alpha\beta} = \begin{bmatrix} G_R & G_K \\ 0 & G_A \end{bmatrix} \tag{21}$$

for fermions and

$$D_{\rho\lambda} = \begin{bmatrix} D_K & D_R \\ D_A & 0 \end{bmatrix} \tag{22}$$

for bosons. Thus in a self-consistent theory we expect $G_{21} = D_{22} = 0$ after a Keldysh rotation, which we will self-consistently check in the derivation process. Green's functions $G_R$ and $D_R$ are the retarded Green's functions, and $G_A$ and $D_A$ are advanced Green's functions, and $G_K$ and $D_K$ are Keldysh Green's functions.

With the structures in Eqs. (21) and (22), the explicit expressions for the components of bosonic self-energies can be written as

$$\Pi_R = \Pi_{21} = -i\frac{g'^2 c^2}{2}\delta(\mathbf{r} - \mathbf{r}') \left[ G_R(x,x')G_K(x',x) + G_K(x,x')G_A(x',x) \right] \tag{23}$$

$$\Pi_A = \Pi_{12} = -i\frac{g'^2 c^2}{2}\delta(\mathbf{r} - \mathbf{r}') \left[ G_A(x,x')G_K(x',x) + G_K(x,x')G_R(x',x) \right] \tag{24}$$

$$\Pi_K = \Pi_{22} = -i\frac{g'^2 c^2}{2}\delta(\mathbf{r} - \mathbf{r}') \left[ G_K(x,x')G_K(x',x) + G_A(x,x')G_R(x',x) + G_R(x,x')G_A(x',x) \right] \tag{25}$$

$$\Pi_{11} = -i\frac{g'^2 c^2}{2}\delta(\mathbf{r} - \mathbf{r}') \left[ G_A(x,x')G_A(x',x) + G_R(x,x')G_R(x',x) \right] \tag{26}$$

The corresponding fermionic self-energies are

$$\Sigma_{R,g'} = \Sigma_{11,g'} = i\frac{g'^2 c^2}{4}\delta(\mathbf{r} - \mathbf{r}') \left[ (D_K(x,x') + D_K(x',x))G_R(x,x') + (D_R(x,x') + D_A(x',x))G_K(x,x') \right] \tag{27}$$

$$\Sigma_{A,g'} = \Sigma_{22,g'} = i\frac{g'^2 c^2}{4}\delta(\mathbf{r} - \mathbf{r}') \left[ (D_K(x,x') + D_K(x',x))G_A(x,x') + (D_A(x,x') + D_R(x',x))G_K(x,x') \right] \tag{28}$$

$$\Sigma_{K,g'} = \Sigma_{12,g'} = i\frac{g'^2 c^2}{4}\delta(\mathbf{r} - \mathbf{r}') \left[ (D_K(x,x') + D_K(x',x))G_K(x,x') + \right. \tag{29}$$

$$\left. + (D_R(x,x') + D_A(x',x)G_R(x,x')) + (D_A(x,x') + D_R(x',x))G_A(x,x') \right] \tag{30}$$

$$\Sigma_{21,g'} = i\frac{g'^2 c^2}{4} \left[ (D_R(x,x') + D_A(x',x))G_A(x,x') + (D_A(x,x') + D_R(x',x))G_R(x,x') \right] \tag{31}$$

If the theory is self-consistent, we should see $\Pi_{11} = \Sigma_{21} = 0$ in the process of evaluating the expressions in Eqs. (23) - (31). The resulting equations are the self-consistent equations for the large-$N$ expansion of the SYK model and can be used as ordinary input into the Keldysh field theory methods.

## II. DERIVATION OF KINETIC EQUATION

Before constructing kinetic equations itself, we first for the sake of completeness define and showcase the basic useful properties of the Wigner transform. We are interested in studying the quasi-classical limit of the theory, and therefore, instead of studying functions $A(x,x')$, we would like to focus on the dependence on the "center of mass"

coordinate $x + x'$. We assume that all the 2-point functions in the theory are slow-varying with the change of $(x+x')/2$ and fast-varying with the change of $(x - x')/2$. It is useful to define a Fourier-transformed in the $x - x'$ degree of freedom quantity as

$$A(x, p) = \int d^3\tilde{x}\, e^{-ip\cdot\tilde{x}} A\left(x + \frac{\tilde{x}}{2}, x - \frac{\tilde{x}}{2}\right) \tag{32}$$

and study it instead of an original 2-point function. In the equation above $p$ is a shorthand notation for a pair $p = (\omega, \mathbf{k})$ similar to $x = (t, \mathbf{r})$, and $p \cdot \tilde{x} = -\omega\tilde{t} + \mathbf{k}\tilde{\mathbf{r}}$. During the derivation of equations that govern Keldysh Green's functions and self-energies we have encountered two operations: one of them is a product of two functions and another is a convolution. The product of two bilocal functions $A(x, x')$ and $B(x, x')$, when Wigner transformed, leads to equations similar to the loop integrals in ordinary quantum field theory:

$$[C(x, x')](x, \Omega, \mathbf{k}) \equiv [A(x, x')B(x'x)](x, \Omega, \mathbf{k}) = \int \frac{d^3 p_B}{(2\pi)^3} A(x, \Omega + \Omega_B, \mathbf{k} + \mathbf{k}_B) B(x, \Omega_B, \mathbf{k}_B), \tag{33}$$

$$[C(x, x')](x, \Omega, \mathbf{k}) \equiv [A(x, x')B(x, x')](x, \Omega, \mathbf{k}) = \int \frac{d^3 p_B}{(2\pi)^3} A(x, \Omega - \Omega_B, \mathbf{k} - \mathbf{k}_B) B(x, \Omega_B, \mathbf{k}_B). \tag{34}$$

In the equations above the rectangular brackets denote the Wigner transform, and $p_B = (\Omega_B, \mathbf{k}_B)$. The second operation in real space, the convolution, can be formally defined as

$$C(x, x') \equiv (A \circ B)(x, x') = \int d^3 y\, A(x, y)B(y, x'). \tag{35}$$

In Section I of Supplemental Information we implicitly encounter this operation in Eqs. (13) and (14), because the inverse operation in those equations is formally defined as an inverse with respect to the convolution defined in Eq. (35), similar to ordinary field theory in real space. The Wigner transform of the convolution, in fact, involves infinite series, however in the quasi-classical limit the series can be truncated to

$$(A \circ B)(x, p) \approx A(x, p)B(x, p) + \frac{i}{2}\left(\frac{\partial A}{\partial x}\frac{\partial B}{\partial p} - \frac{\partial A}{\partial p}\frac{\partial B}{\partial x}\right) \tag{36}$$

Eq. (36) is the central approximation of the theory that limits the applicability of the theory to the large wavelength and small frequency perturbations, since it involves the infinite series truncation under assumption that the higher order terms in the series are small.

In this paper we are interested in non-linear conductivity, and thus a current density expectation value $\mathbf{j}$. In Keldysh field theory every physical observable is obtained by the operator insertions on the forward in time contour, thus with the "+" operators. The expectation for the current $j$ can be written as

$$\langle \hat{j}(t, \mathbf{r})\rangle \equiv \langle \hat{j}_{++}(t, \mathbf{r})\rangle = \frac{e}{2imN}\sum_i \left[\hat{\psi}_{i+}^\dagger(t, \mathbf{r})\nabla\hat{\psi}_{i+}(t, \mathbf{r}) - \nabla\hat{\psi}_{i+}^\dagger(t, \mathbf{r})\hat{\psi}_{i+}(t, \mathbf{r})\right], \tag{37}$$

which can be rewritten through a "lesser" Green's function $G^<(x, x') = (G_K(x, x') - G_R(x, x') + G_A(x, x'))/2$

$$\langle \hat{j}(t, \mathbf{r})\rangle = -\frac{e}{2m}\lim_{\substack{t' \to t \\ \mathbf{r}' \to \mathbf{r}}}[\nabla_\mathbf{r} G^<(x, x') - \nabla_{\mathbf{r}'} G^<(x, x')], \tag{38}$$

instead of $G_{++}$, since $G_{++}(x, x') = \theta(t - t')G^>(x, x') + \theta(t' - t)G^<(x, x')$.

if the "++" component of the Green's function is written through the degrees of freedom in Keldysh basis. Performing the Wigner transform leads to

$$\langle \hat{j}(t, \mathbf{r})\rangle = -ie \int \frac{d\omega}{2\pi}\int \frac{d^2 k}{(2\pi)^2}\mathbf{v}_F G^<(t, \mathbf{r}, \omega, \mathbf{k}). \tag{39}$$

To proceed, forward from here, we would like to establish the intuition for the choice of function $f(x, \omega, \theta)$ from the main text in more detail. In general, the thermal equilibrium expressions for a lesser Green's functions $G^<$ and $D^<$ according to Keldysh field theory are completely fixed by the structure of $G_R$ and $D_R$ correspondingly and are given by

$$G^<(\omega, \mathbf{k}) = -\bar{f}(\omega)(G_R(\omega, \mathbf{k}) - G_A(\omega, \mathbf{k})), \qquad \bar{f}(\omega) = \frac{1}{e^{\beta\omega} + 1}, \tag{40}$$

$$D^<(\Omega, \mathbf{q}) = \bar{f}_B(\Omega)(D_R(\Omega, \mathbf{q}) - D_A(\Omega, \mathbf{q})), \qquad \bar{f}_B(\Omega) = \frac{1}{e^{\beta\Omega} - 1}. \tag{41}$$

We now show that a closed system of kinetic equations can be obtained for these functions. According to [5, 6], it is possible to define a quantity in Keldysh field theory that would play a role of an analog of occupation number in the free theory even when the notion of the quasiparticle is suppressed by damping. The conditions for the definition consistent being: self-energies independent from momentum and presence of a sharp Fermi surface. The presence of a sharp Fermi-surface is required for a presence of a relatively sharp spectral density peak, while the independence of self-energies of momentum is required to make the system of kinetic equations being closed under the new "occupation numbers".

Let us start with the equilibrium Green's functions. The retarded and advanced ones, present in (40) can be obtained from the calculations in Matsubara field theory in [1–3] by analytic continuation. Equilibrium retarded Green's functions takes a form

$$\bar{G}_R(\omega, \mathbf{k}) = \frac{1}{\omega - v_F k - \mathbf{v}_F \cdot \mathbf{A} - \Sigma_R(\omega)}, \qquad \bar{D}_R(\Omega, \mathbf{q}) = \frac{1}{\Omega^2 - c^2 q^2 - m^2 + i c_d \Omega}. \tag{42}$$

The parameter $v_F$ is a Fermi velocity, $\Sigma_R(\omega)$ - momentum independent self-energy rougly independent of $\omega$ at small $\omega$, $c \lesssim v_F$ - boson velocity, $m$ - boson thermal mass, and $c_d = g'^2 c^2 k_F^2 / 4\pi v_F^2$ - self-consistent Landau damping. The expressions above assumes a linearized dispersion of the fermion band in the vicinity of the Fermi surface. Interestingly, in this model even away from equilibrium the self-energies for boson and fermion don't depend on the momentum, which allows us to define

$$f(x, \omega, \theta_{\mathbf{k}}) = \frac{\int_{-\infty}^{+\infty} \frac{k_F dk}{2\pi} G^<(x, \omega, \mathbf{k})}{-2i \int_{-\infty}^{+\infty} \frac{k_F dk}{2\pi} \operatorname{Im} G_R(x, \omega, \mathbf{k})}, \qquad f_B(x, \Omega, \theta_{\mathbf{q}}) = \frac{\int_0^{+\infty} \frac{c^2 q dq}{2\pi} D^<(x, \Omega, \mathbf{q})}{2i \int_0^{+\infty} \frac{c^2 q dq}{2\pi} \operatorname{Im} D_R(x, \Omega, \mathbf{q})}, \tag{43}$$

where $\theta_{\mathbf{k}}$ and $\theta_{\mathbf{q}}$ are the angles that define the direction of $\mathbf{k}$ and $\mathbf{q}$ correspondingly. Substituting Eqs. (42) into the definition of $f$ and $f_B$ leads us to

$$f(x, \omega, \theta_{\mathbf{k}}) = -i \frac{v_F}{k_F} \int \frac{k_F dk}{(2\pi)} G^<(x, \omega, \mathbf{k}), \qquad f_B(x, \Omega, \theta) = \frac{i}{2} \frac{1}{B(x, \Omega, \theta_{\mathbf{q}})} \int_0^{+\infty} \frac{c^2 q dq}{2\pi} D^<(x, \Omega, \mathbf{q}), \tag{44}$$

with $B(\Omega)$ being proportional to the momentum integrated spectral density

$$B(x, \Omega, \theta_{\mathbf{q}}) = -\int \frac{c^2 q dq}{2\pi} \operatorname{Im} D_R(x, \Omega, q). \tag{45}$$

From the expressions above we can instantly see the importance of the sharp Fermi surface requirement: the generalized fermionic distribution $f$ is just proportional to the integral of a lesser Green's function $G^<$ with a coefficient that is just a density of states at the Fermi surface regardless of the perturbation. Meanwhile, the generalized bosonic distribution function $f_B$ has a complicated function modifying the integral of $D^<$. Below we will show that $D_R$ stays unperturbed from equilibrium value at least in perturbation theory, thus $B$ is independent from $x$ and $\theta_{\mathbf{q}}$. However, the non-trivial dependence on $\Omega$ will remain the feature of the theory. Note that we only used Eqs. (40) and (41) as an inspiration for constructing $f$ and $f_B$ in such a way that they coincide with bosonic and fermionic equilibrium distributions when the system is in thermal equilibrium. As a consistency check of the theory, this fact will be explicitly proven below. Therefore, in all the calculation below we will treat $\bar{f}(\omega)$ and $\bar{f}_B(\Omega)$ as equilibrium expressions for $f$ and $f_B$ without assuming any explicit form.

It is useful to re-express the fermionic and bosonic non-equilibrium self-energy expressions obtained in Eqs. (23) - (31) with the use of $f$ and $f_B$. We start from bosonic self-energies $\Pi_R$, $\Pi_A$, $\Pi_K$, and $\Pi_{11}$. After performing a Wigner transform to Eqs. (23) - (26) and substituting Eq. (44) into the result we obtain

$$\Pi_R(x, \omega) = -i \frac{g'^2 k_F^2}{v_F^2} \int_{-\infty}^{+\infty} \frac{d\omega}{2\pi} \int_0^{2\pi} \frac{d\theta}{2\pi} \left[ f(x, \omega, \theta) - f(x, \omega + \Omega, \theta) \right] \tag{46}$$

$$\Pi_A(x, \omega) = i \frac{g'^2 k_F^2}{v_F^2} \int_{-\infty}^{+\infty} \frac{d\omega}{2\pi} \int_0^{2\pi} \frac{d\theta}{2\pi} \left[ f(x, \omega, \theta) - f(x, \omega + \Omega, \theta) \right] \tag{47}$$

$$\Pi_K(x, \omega) = i \frac{g'^2 k_F^2}{2v_F^2} \int_{-\infty}^{+\infty} \frac{d\omega}{2\pi} \int_0^{2\pi} \frac{d\theta}{2\pi} \frac{d\theta'}{2\pi} \left[ (1 - 2f(x, \omega + \Omega, \theta))(1 - 2f(x, \omega, \theta')) - 1 \right]. \tag{48}$$

The expression for $\Pi_K$ has two momentum angles $\theta$ and $\theta'$ decoupled due to spatially randomized coupling, which averages out the interaction of all but density harmonics. The expression for $\Pi_{11}$ reads

$$\Pi_{11}(x, \Omega) = -i \frac{g'^2 c^2}{2} \int \frac{d^2 k}{(2\pi)^2} \frac{d^2 k'}{(2\pi)^2} \int_{-\infty}^{+\infty} \frac{d\omega}{2\pi} [G_A(x, \omega + \Omega, \mathbf{k}) G_A(x, \omega, \mathbf{k}') + G_R(x, \omega + \Omega, \mathbf{k}) G_R(x, \omega, \mathbf{k}')] = 0, \tag{49}$$

because the poles of the products of two retarded and two advanced Green's functions have poles only on one side of the contour. Therefore, correct bosonic field causality structure is preserved and expressions for bosonic field are so far self-consistent.

Before considering fermionic self-energies, we will detour to understand the structure of the bosonic self energies better. First of all, we expect the total number of fermions $N_F$ to be unchanged by external perturbations. The total fermion number can be expressed as

$$N_F = \frac{1}{N} \sum_i \int d^2 \mathbf{r} \langle \psi_{i+}^\dagger(t, \mathbf{r}) \psi_{i+}(t, \mathbf{r}) \rangle = \int d^2 \mathbf{r} \int_{-\infty}^{+\infty} \frac{d\omega}{2\pi} \int_0^{2\pi} \frac{d\theta}{2\pi} f(x, \omega, \theta). \tag{50}$$

Since in this work we resort only to spatially homogeneous fields, appearance of the charge redistribution in space is not expected and function $f$ is independent of $\mathbf{r}$. Since the charged particle density remains constant in space, one can express charge conservation as

$$\int_{-\infty}^{+\infty} \frac{d\omega}{2\pi} \int_0^{2\pi} \frac{d\theta}{2\pi} (f(x, \omega, \theta) - \bar{f}(\omega)) = 0 \tag{51}$$

Thus, expression for $\Pi_R(t, \Omega)$ will be equal to the equilibrium expression $\bar{\Pi}_R(\Omega)$, because using Eq. (51) it can be rewritten as as

$$\Pi_R(x, \omega) = -i \frac{g'^2 k_F^2}{v_F^2} \int_{-\infty}^{+\infty} \frac{d\omega}{2\pi} \left[ \bar{f}(\omega) - \bar{f}(\omega + \Omega) \right] = \bar{\Pi}_R(\Omega) = -i c_d \Omega, \tag{52}$$

where $\bar{\Pi}_R$ is obtained by analytic continuation from thermal field theory in previous works. Assuming $\bar{f}(\omega) = 1/(e^{\beta\omega} + 1)$, the value of the integral is consistent with the thermal field theory expression for $c_d = g'^2 c^2 k_F^2 / 4\pi v_F^2$. Since bosonic self-energy remains uncorrected by non-equilibrium effects, the thermal mass $m^2$ and retarded bosonic Green's function $D_R$ will also be equal to their equilibrium expressions. Thus

$$D_R(t, \Omega, \mathbf{q}) = \bar{D}_R(\Omega, \mathbf{q}) = \frac{1}{\Omega^2 - c^2 q^2 + m^2 + i c_d \Omega}, \quad B(t, \Omega) = B(\Omega) = -\int \frac{c^2 q dq}{2\pi} \operatorname{Im} \bar{D}_R(\Omega, \mathbf{q}) = \frac{1}{4\pi} \arctan\left(\frac{c_d \Omega}{m^2}\right). \tag{53}$$

Analogous but complex conjugated expression one can obtain for $\bar{\Pi}_A(\Omega)$. By applying Wigner transform to the expressions for fermionic self-energies $\Sigma_R$, $\Sigma_A$, and $\Sigma_K$ given by Eqs. (27) - (31) and substituting expressions for $f$ and $f_B$ from Eq. (44) we obtain

$$\operatorname{Im} \Sigma_R(x, \omega) = -\frac{\Gamma}{2} - \frac{g'^2 k_F}{v_F} \int \frac{d\Omega}{2\pi} \int \frac{d\theta}{2\pi} B(\Omega) \left[ f_B(x, \Omega, \theta) - f(x, \omega - \Omega, \theta) + 1 \right] \tag{54}$$

$$\operatorname{Re} \Sigma_R(x, \omega) = \frac{g'^2 k_F}{2 v_F} \int \frac{d\Omega}{2\pi} \int \frac{d\theta}{2\pi} B'(\Omega) (2 f(x, \omega - \Omega, \theta) - 1), \tag{55}$$

$$\Sigma_K(x, \omega) = 4 i \Gamma \int \frac{d\theta}{2\pi} f(x, \omega, \theta) - 2 i \Gamma + i \frac{g'^2 k_F}{v_F} \int \frac{d\Omega}{2\pi} \int \frac{d\theta d\theta'}{(2\pi)^2} B(\Omega) (2 f_B(\Omega, \theta') + 1)(f(\omega + \Omega, \theta) + f(\omega - \Omega, \theta) - 1) \tag{56}$$

where $\Gamma = v^2 k_F / v_F$ and

$$B'(\Omega) = c^2 \int \frac{d^2 q}{(2\pi)^2} \operatorname{Re} D_R(\Omega, \mathbf{q}) = \frac{1}{8\pi} \ln\left(\frac{c^4 \Lambda_q^4}{m^4 + c_d^2 \Omega^2}\right) \tag{57}$$

with $\Lambda_q$ being a UV cutoff such that $c\Lambda_q \sim T_F$. The expression for $\Sigma_{21}$ is

$$\Sigma_{21, g'} = i \frac{g'^2 c^2}{4} \int \frac{d^2 k}{(2\pi)^2} \frac{d^2 k'}{(2\pi)^2} \int \frac{d\Omega}{2\pi} [D_R(x, \Omega, \mathbf{q}) G_A(x, \omega - \Omega, \mathbf{k}) + D_A(x, \Omega, \mathbf{q}) G_A(x, \omega + \Omega, \mathbf{k}) +$$
$$+ D_A(x, \Omega, \mathbf{q}) G_R(x, \omega - \Omega, \mathbf{k}) + D_R(x, \Omega, \mathbf{q}) G_R(x, \omega + \Omega, \mathbf{k})] = 0 \tag{58}$$

due to all the poles for contour integration over $\Omega$ being on one side of the contour for every term in the integral. Therefore, the causality structure of the fermionic Green's functions is also self-consistent, and our Keldysh expansion is performed correctly.

Now we proceed to deriving the closed form of the kinetic equations for $f$ and $f_B$ with the use of all the self-energies expressed as functionals of $f$ and $f_B$. To perform the derivation, we need to transform the Dyson equations for $G_K$ and $D_K$, which according to Eqs. (13) and (14) are

$$G_K = G_R \circ \Sigma_K \circ G_A, \qquad D_K = D_R \circ \Pi_K \circ D_A, \tag{59}$$

can be converted into the corresponding Dyson equations for $G^<$ and $D^<$:

$$G^< = \frac{1}{2} G_R \circ \Sigma_K \circ G_A - \frac{1}{2} G_R + \frac{1}{2} G_A, \qquad D^< = \frac{1}{2} D_R \circ \Pi_K \circ D_A - \frac{1}{2} D_R + \frac{1}{2} D_A. \tag{60}$$

To reduce these Dyson equations to kinetic equations, we follow approach in Ref. [4]. As the next step towards kinetic equations we compute the commutator of inverse bare Green's functions with the lesser Green's functions, with the use of Eq. (60):

$$[G_0^{-1}; G^<] = \Sigma_R \circ G^< - G^< \circ \Sigma_A + \frac{1}{2} \left( \Sigma_K + \Sigma_A - \Sigma_R \right) \circ G_A - \frac{1}{2} G_R \circ \left( \Sigma_K + \Sigma_A - \Sigma_R \right) \tag{61}$$

$$[D_0^{-1}; D^<] = \Pi_R \circ D^< - D^< \circ \Pi_A + \frac{1}{2} \left( \Pi_K + \Pi_A - \Pi_R \right) \circ D_A - \frac{1}{2} D_R \circ \left( \Pi_K + \Pi_A - \Pi_R \right) \tag{62}$$

Eqs. (61) and (62) are the kinetic equations for the lesser functions in the real space. To show that they are, in fact, closed equations for $f$ and $f_B$, we perform a Wigner transform on Eqs. (61) and (62) and integrate them over corresponding absolute values of momenta. The independence of self-energies on momenta will allow to close the equations on momenta-independent $f$ and $f_B$. Before obtaining full, non-equilibrium, expressions, it is useful to first understand the relations between the equilibrium solutions to these equations. In thermal equilibrium Eqs. (61) and (62) reduce to

$$\left( 2 \bar{f}(\omega) - 1 \right) \operatorname{Im} \bar{\Sigma}_R(\omega) - \frac{i}{2} \bar{\Sigma}_K(\omega) = 0, \tag{63}$$

$$\left( \bar{f}_B(\Omega) + \frac{1}{2} \right) \operatorname{Im} \bar{\Pi}_R(\Omega) + \frac{i}{4} \bar{\Pi}_K(\Omega) = 0. \tag{64}$$

These equations can be trivially satisfied by a solution of the form

$$\bar{f}(\omega) = \frac{1}{e^{\beta \omega} + 1}, \qquad \bar{f}_B(\Omega) = \frac{1}{e^{\beta \Omega} - 1}, \qquad \beta = \frac{1}{T}. \tag{65}$$

Therefore, our initial guess about the structure of the thermal equilibrium ground state is correct. After a Wigner transform and integration over the absolute value of momenta, with the use of relations in Eqs. (63) we obtain the equations for fermion and boson distributions. Kinetic equation for fermions can be the most conveniently written as

$$A[f, \partial_t f] + e(\mathbf{v}_F \cdot \mathbf{E}) \partial_\omega f = -\Gamma \left( f - f_0 \right) - \bar{g}(\omega, T)(f - \bar{f}) + I_{g'}[f_B, f] - (f - 1/2) \delta g[f_B, f], \tag{66}$$

where $f_0 = \int \frac{d\theta}{2\pi} f$, kinetic term $A[f, \partial_t f]$ can be written as

$$A[f, \partial_t f] = (1 - \partial_\omega \operatorname{Re} \Sigma_R) \partial_t f + \partial_\omega f \partial_t \operatorname{Re} \Sigma_R \tag{67}$$

with $\operatorname{Re} \Sigma_R$ given by Eq. (55), collision integral term $g(\omega, T) = - \operatorname{Im} \Sigma_{R, g'}$, and terms $I_{g'}$ nad $I_{\delta g}$ are

$$I_{g'}[f_B, f] = \frac{g'^2 k_F}{2 v_F} \int \frac{d\Omega}{2\pi} \int \frac{d\theta d\theta'}{(2\pi)^2} (K_{g'}[f_B, f] - K_{g'}[\bar{f}_B, \bar{f}]) \tag{68}$$

$$\delta g[f_B, f] = \frac{2 g'^2 k_F}{v_F} \int \frac{d\Omega}{2\pi} \int \frac{d\theta}{2\pi} (K_{\delta g}[f_B, f] - K_{\delta g}[\bar{f}_B, \bar{f}]) \tag{69}$$

with

$$K_{g'}[f_B, f] = B(\Omega)(2 f_B(\Omega, \theta) + 1)(f(\omega + \Omega, \theta') + f(\omega - \Omega, \theta') - 1) \tag{70}$$

$$K_{\delta g}[f_B, f] = B(\Omega)(f_B(\Omega, \theta) - f(\omega - \Omega, \theta)) \tag{71}$$

Kinetic equation for boson can be written as

$$\partial_t f_B + c_d (f_B - \bar{f}_B) = I_B[f] + C(\Omega) \partial_t I_B[f], \tag{72}$$

where we introduced following notation:

$$I_B[f] = -\frac{c_d}{4\Omega} \int_{-\infty}^{+\infty} d\omega \int_0^{2\pi} \frac{d\theta}{2\pi} \frac{d\theta'}{2\pi} (K_B[f] - K_B[\bar{f}]), \qquad K_B[f] = (1 - 2f(x, \omega + \Omega, \theta))(1 - 2f(x, \omega, \theta')), \tag{73}$$

and $C(\Omega) = \partial_\Omega B'(\Omega)/2B(\Omega)$. Expression for $c_d = g'^2 k_F^2 c^2 / 4\pi v_F^2$ coincides with the corresponding thermal field theory expression.

Kinetic equations in Eqs. (66) and (72) have a very complicated form, however, as we will see below, some of the terms will always remain small and thus can be neglected leading to a more tractable model. First of all, we would like to note that equations above use the "small external frequency assumption" used in Eq. (36), which requires inequalities

$$|\partial_\omega \operatorname{Re} \bar{\Sigma}_R \partial_t \delta f| \ll |\operatorname{Im} \bar{\Sigma}_R f| \tag{74}$$

$$|\partial_\omega \bar{f} \partial_t \operatorname{Re} \delta \Sigma_R| \ll |(2\bar{f} - 1) \operatorname{Im} \bar{\Sigma}_R| \tag{75}$$

$$|C(\Omega) \partial_t \Pi_K| \ll |\Pi_K| \tag{76}$$

to hold. All three of these inequalities can be eventually simlified to $\nu \leq T$, where $\nu$ is a characteristic frequency of external perturbation and $T$ is the temperature of the system. Additionally, we require $T \leq c_d$ to hold, so Landau damping term $c_D \Omega \gg \Omega^2$ in the structure of $D_R$. These inequalities naturally create a hierarchy of energy scales

$$\nu \lesssim T \sim \omega \sim \Omega \ll c_d \lesssim T_F \tag{77}$$

that allows us to simplify the kinetic equations. We start our analysis from the boson kinetic equation. We notice that all the terms in Eq. (72) are of the order of $c_d f_B$, while the $\partial_t f_B$ term is of the order of $\nu f_B$. The term proportional to $C(\Omega)$ can be neglected for a similar reason as one proportional to $c_d \nu / T f_B$. Thus the time derivative term in Eq. (72) can be neglected. We will introduce one more modification, which this time is phenomenological. In a more realistic system one can imagine other types of disorder and interaction that might lead to an additional Landau damping on top of the self-consisted damping that comes from the Fermion interaction. Thus the damping term $c_d'(f_B - \bar{f}_B)$ can have a constant $c_d' \geq c_d$. Thus, one of the ways to phenomenologically include extra damping is to modify the constant $c_d$ in the landau damping term. We define a coeficient $\lambda = c_d/c_d'$ and we can write down a resulting boson kinetic equation as

$$f_B(t, \Omega) = \bar{f}_B(\Omega) + \lambda I_B[f](t, \Omega). \tag{78}$$

Thus, when no additional Landau damping is present ($\lambda = 1$), the non-equilibrium distribution of a fermion drives the distribution of a boson out of equilibrium. When the additional Landau damping dominates over self-consistent part $c_d$ ($\lambda = 0$), the boson will always remain in a thermal distribution. Thus case of $\lambda = 0$ corresponds to a scenario considered in [6], while $\lambda \neq 0$ is new to this model and interpolates betweeen the thermalized boson and a fully self-consistent boson dynamics.

We begin simplifying the fermion kinetic equation from considering the structure of equilibrium fermionic self-energies as a function of $T$, $\omega$, and $m$. We consider a small $T$ expression for the fermion self-energy derived in [3, 6] and given by

$$\bar{\Sigma}_R = -i\Gamma + i \frac{g'^2 k_F}{8\pi v_F} T \left[ \ln\left(\frac{c^2 \Lambda_q^2}{m^2}\right) - 2\left( \ln\Gamma\left(\frac{c^2 \Lambda_q^2}{2\pi c_d T} + \frac{1}{2} - \frac{i\omega}{2\pi T}\right) - \ln\Gamma\left(\frac{m^2}{2\pi c_d T} + \frac{1}{2} - \frac{i\omega}{2\pi T}\right) \right. \right.$$
$$\left. \left. - \ln\Gamma\left(\frac{c^2 \Lambda_q^2}{2\pi c_d T}\right) + \ln\Gamma\left(\frac{m^2}{2\pi c_d T}\right) \right) \right], \tag{79}$$

where $\Gamma(z)$ is an Euler gamma-function and $\Lambda_q$ – a UV cutoff of the boson. Assuming that scale of the UV cutoff $\Lambda_q$ is dominant over all the other scales, the expression for the self-energy can be further simplified based on the structure of $m(T)$. The structure of boson thermal mass at the critical value, according to [3], in the leading order is

$$m^2(T) \approx \frac{\pi c_d T}{\ln\left(\frac{\Lambda_q^2 c^2}{c_d T}\right)} \tag{80}$$

With this expression for thermal mass, relevant for the equation quantities become (expanding for $c\Lambda_q \gg m, c_d, T$):

$$a(\omega, T) = 1 - \partial_\omega \operatorname{Re} \bar{\Sigma}_R = 1 + \frac{g'^2 k_F}{8\pi v_F} \left[ \ln\left(\frac{c^2 \Lambda_q^2}{2\pi c_d T}\right) - \operatorname{Re} \psi^{(0)}\left(\frac{1}{2} + \frac{m^2}{2\pi c_d T} - \frac{i\omega}{2\pi T}\right) \right] \tag{81}$$

$$\bar{g}(\omega, T) = -2\operatorname{Im}\left[\bar{\Sigma}_R + i\Gamma\right] = \frac{g'^2 k_F}{4\pi v_F} T \left[ -\ln\left(\frac{2\pi c_d T}{m^2}\right) + 2\ln\Gamma\left(\frac{m^2}{2\pi c_d T}\right) - 2\operatorname{Re}\ln\Gamma\left(\frac{1}{2} + \frac{m^2}{2\pi c_d T} - \frac{i\omega}{2\pi T}\right) \right] \tag{82}$$

For $T \ll T_F$, $m^2 \ll c_d T$ and the expressions above in the leading order (assuming $\omega$ to be of the order or smaller than $T$) become

$$a_{\text{cr}}(\omega, T) \approx 1 + \frac{g'^2 k_F}{8\pi^2 v_F} \left[ \ln\left( \frac{c^2 \Lambda_q^2}{2\pi c_d T} \right) \right] \tag{83}$$

$$g_{\text{cr}}(\omega, T) \approx \frac{g'^2 k_F}{4\pi v_F} T \ln\left( \frac{2 c_d T}{m^2} \text{ch}\left( \frac{\omega}{2T} \right) \right) \tag{84}$$

The structure of the self-energy $\bar{\Sigma}_R$ sufficiently away from the critical doping, is still guided by Eq. (79), but now involves

$$m^2(T) \approx \Delta^2 = \text{const}, \tag{85}$$

which gives the boson a temperature-independent gap $\Delta$ in the leading order. Assuming we are in the low temperature regime and $\Delta \gg T$, the expressions for $a(\omega, T)$ and $g(\omega, T)$ simplify to the form one would expect from a Fermi liquid:

$$a_{\text{FL}}(\omega, T) = 1 + \frac{g'^2 k_F}{8\pi^2 v_F} \ln\left( \frac{c^2 \Lambda_q^2}{e \Delta^2} \right), \tag{86}$$

$$g_{\text{FL}}(\omega, T) = \frac{g'^2 k_F c_d}{8\pi^2 v_F \Delta^2} \left( \pi^2 T^2 + \omega^2 \right). \tag{87}$$

The expressions above allow us to better understand the structure of the kinetic and collision integral terms in Eq. (66) close and away from the critical point. From the expression in Eq. (67) we can see that in the critical case the second term, proportional to $\partial_t \Sigma_R$, is suppressed by $\nu / T \ln(T_F / T)$ in comparison to the first term. Away from the critical point, the second term is suppressed by $\nu T / \Delta^2$. Thus in both cases the second term can be neglected. Moreover, the non-equilibrium correction to $\partial_\omega \text{Re} \Sigma_R$ can also be neglected in the first term in Eq. (67), because in the perturbation theory it will always produce terms suppressed by similar to the aforementioned second term factors. Therefore, the complicated dynamic term can be reduced to

$$A[f, \partial_t f] = a(\omega, T) \partial_t f(t, \omega, \theta), \tag{88}$$

where $a(\omega, T)$ takes a form $a_{\text{cr}}$ close to the QCP, or $a_{\text{FL}}$ away from the QCP with

$$a_{\text{cr}}(\omega, T) = 1 + \frac{g'^2 k_F}{8\pi^2 v_F} \ln\left( \frac{c^2 \Lambda_q^2}{2\pi c_d T} \right), \tag{89}$$

$$a_{\text{Fl}}(\omega, T) = 1 + \frac{g'^2 k_F}{8\pi^2 v_F} \ln\left( \frac{c^2 \Lambda_q^2}{e \Delta^2} \right). \tag{90}$$

To summarize the structure of the theory, it is convenient to write it in terms of the angular harmonics $e^{i\theta m}$

$$f_m(t, \omega) = \int_0^{2\pi} \frac{d\theta}{2\pi} e^{-i\theta m} f(t, \omega, \theta), \qquad f_{Bm}(t, \omega) = \int_0^{2\pi} \frac{d\theta}{2\pi} e^{-i\theta m} f_B(t, \Omega, \theta). \tag{91}$$

The equations for $f_m$ with $m \neq 0$ can be written in a relatively simple form

$$\left( a(\omega, T) \partial_t + g(t, \omega, T) + \Gamma_m \right) f_m = -\frac{e v_F}{2} \partial_\omega (\mathcal{E}^* f_{m-1} + \mathcal{E} f_{m+1}), \tag{92}$$

where $\mathcal{E} = E_x + i E_y$. In our model all $\Gamma_m = \Gamma$ for $m \neq 0$, but we would consider a more generic model with different eigenvalues for different harmonics. Since disorder is elastic, we naturally get $\Gamma_0 = 0$ exactly. The equation for an $m = 0$ harmonic of $f_m$ is more complicated:

$$a(\omega, T) \partial_t f_0 + \bar{g}(\omega, T)(f_0 - \bar{f}) + (f_0 - 1/2) \delta g(t, \omega, T) - I_{g'}[f_0, f_{B0}] = -\frac{e v_F}{2} \partial_\omega (\mathcal{E}^* f_{-1} + \mathcal{E} f_{+1}) \tag{93}$$

The and $a(\omega, T)$ and $g(t, \omega, T)$ for the form close to the critical point take a form $a_{\text{cr}}(t, \omega, T)$ $g_{\text{cr}}(t, \omega, T)$, and $a_{\text{FL}}(t, \omega, T)$ and $g_{\text{FL}}(t, \omega, T)$ away from the critical point. The expressions for $a_{\text{cr}}(\omega, T)$ and $a_{\text{FL}}(\omega, T)$ are given by Eqs. (89) and (90) correspondingly. Expressions for $g_{\text{cr}}(t, \omega, T)$ and $g_{\text{Fl}}(r, \omega, T)$ can be written as

$$g_{\text{cr}}(t, \omega, T) = \bar{g}_{\text{cr}}(\omega, T) + \delta g[f_0, f_{B0}], \qquad g_{\text{FL}}(t, \omega, T) = \bar{g}_{\text{FL}}(\omega, T) + \delta g[f_0, f_{B0}]. \tag{94}$$

Expressions for $\bar{g}_{\rm cr}(\omega, T)$ and $\bar{g}_{\rm FL}(\omega, T)$ are given by Eqs. (84) and (87), while $I_{g'}$ and $\delta g$ is given be Eqs. (68) and (69) with an appropriate choice of mass $m^2$.

As a last note, we would like to point out a few features of the behavior of $f_0$ collision integral. For fully consistent boson dynamics with $\lambda = 1$, the linearized collision integrals for density harmonics $f_0$ and $f_{B0}$ have a natural 0-eigenvalue mode

$$\delta f_0 \sim \frac{\partial}{\partial T}\bar{f}(\omega), \qquad \delta f_{B0} \sim \frac{\partial}{\partial T}\bar{f}_B(\Omega), \tag{95}$$

since thermal distributions naturally satisfy kinetic equations of this model. These modes are associated with the enery conservation. Probably a more peculiar feature belongs to a $\lambda = 0$ model, where the boson is thermalized. Since the temperature of the thermal bath is fixed by the strong boson thermalization and the bath strongly interacts with the fermions, the energy is no longer conserved. The smallest eigenvalue of the fermionic collision integral now will correspond to

$$\delta f_0 \sim \frac{\partial^2}{\partial \omega^2}f(\omega), \qquad \Gamma_0 = -\frac{g'^2 k_F}{4\pi v_F} \tag{96}$$

with eigenvalue $\Gamma_0$, which can be checked by linearizing $f_0 = \bar{f} + \delta f_0$ in $\delta f$ the expression in (66). Linearized part of the collision integral integral from Eq. (93) is produced by $\bar{g}$, $\delta g$, and $I_{g'}$ terms and reads

$$I_{\rm lin}[\delta f_0](\omega) = -\frac{g'^2 k_F}{v_F}\int \frac{d\Omega}{2\pi}B(\Omega)\left[\left(\coth\left(\frac{\Omega}{2T}\right) + \tanh\left(\frac{\omega-\Omega}{2T}\right)\right)\delta f_0(\omega) + \right.$$
$$\left. + \tanh\left(\frac{\omega}{2T}\right)\delta f_0(\omega-\Omega) - \coth\left(\frac{\Omega}{2T}\right)\delta f_0(\omega-\Omega)\right] \tag{97}$$

When $\delta f_0(\omega) = \partial_\omega^2 \bar{f}(\omega)$ is substituted in the equation above it simplifies to

$$I_{\rm lin}[\partial_\omega^2\bar{f}](\omega) = -\frac{g'^2 k_F}{v_F}\partial_\omega^2\bar{f}\int_0^{+\infty}\frac{d\Omega}{2\pi}B(\Omega)\coth\left(\frac{\omega}{2T}\right)\left[\tanh^2\left(\frac{\omega+\Omega}{2T}\right) - \tanh^2\left(\frac{\omega-\Omega}{2T}\right)\right] \approx -\frac{g'^2 k_F}{4\pi v_F}\partial_\omega^2\bar{f}(\omega). \tag{98}$$

Where the last approximate equality only holds when $m^2 \ll c_d T$ that holds at sufficiently low temperatures at the critical point. In this work we will not consider the case of thermalized boson away from the quantum critical point and will restrict ourselves to only a fully dynamic case.

## III. PERTURBATIVE SOLUTION

### A. Distribution function

We construct the perturbation theory in electric field $\mathbf{E}$ to obtain the higher orders of conductivity. We denote the order of perturbation in $\mathbf{E}$ with a superscript:

$$f_m = \sum_i f_m^{(i)}, \tag{99}$$

where $f_m^{(i)}$ is a contribution of $i$'th order of perturbation theory. We start from the linear perturbation. It involves only harmonics with $m = \pm 1$, which can be extracted from a linearized in $E$ version of Eq. (92) and can be conveniently written with the help of

$$W_m(\nu, \omega, T) = \frac{k_F v_F}{i\nu a(T) + \Gamma_m + \bar{g}(\omega, T)} \tag{100}$$

for $m \neq 0$ assuming the incoming electric field takes a form

$$\mathcal{E} = \sum_i \mathcal{E}_{\nu_i}e^{i\nu_i t} \qquad \mathcal{E}_{\nu_i} = E_{\nu_i x} + iE_{\nu_i y}. \tag{101}$$

The expressions for $f_{\pm 1}^{(1)}$ become

$$f_{-1}^{(1)} = (f_{+1}^{(1)})^* = -\frac{e}{2k_F}\sum_i W_1(\nu_i, \omega, T)\mathcal{E}_{\nu_i}\partial_\omega\bar{f}(\omega). \tag{102}$$

The second order perturbation involves quadrupolar harmonics $m = \pm 2$ and a density harmonic $m = 0$. The quadrupolar harmonics at the level of second order perturbation theory are still described by the linearized version of Eq. (92) and thus can be easily written as

$$f_{-2}^{(2)} = (f_{+2}^{(2)})^* = \frac{e^2}{4k_F^2} \sum_{i,j} W_2(\nu_{ij}, \omega, R) W_1(\nu_i, \omega, T) \mathcal{E}_{\nu_j} \mathcal{E}_{\nu_i} \partial_\omega^2 \bar{f}(\omega), \tag{103}$$

where $\nu_{ij} = \nu_i + \nu_j$. The density harmonic is more complicated due to complicated structure of the collision integral in Eq. (93). In general, it has to be solved separately for every value of $\lambda$ in Eq. (78), but we are going to focus on the case of a thermalized boson $\lambda = 0$ and a fully dynamic boson $\lambda = 1$. Since the structure and behavior of the collision integral is very different in the limiting cases. First consider the case of thermalized boson $\lambda = 0$, it appears that the source term in linearized Eq. (78) is proportional to $\partial_\omega^2 \bar{f}(\omega)$, which is, luckily, an approximate eigenvector of the collision operator. Thus the response for thermalized boson can be conveniently written with a help of

$$W_0(\nu, T) = \frac{k_F v_F}{i\nu a(T) + g_0(T, \lambda)}, \qquad g_0(T, \lambda = 0) = \frac{g'^2 k_F}{4\pi v_F} T \sim g(\omega \sim T, T). \tag{104}$$

The expression for $f_0^{(2)}$ becomes

$$f_{0,\lambda=0}^{(2)} = \frac{e^2}{4k_F^2} \sum_{i,j} W_0(\nu_{ij}, \omega, T) \left[ \mathcal{E}_{-\nu_j}^* \mathcal{E}_{\nu_i} + \mathcal{E}_{\nu_j} \mathcal{E}_{-\nu_i}^* \right] \frac{\partial}{\partial \omega} \left[ W_1(\nu_i, \omega, T) \frac{\partial \bar{f}(\omega)}{\partial \omega} \right] \tag{105}$$

The structure of the response for a dynamic boson ($\lambda = 1$) is principally different. With a dynamic boson the model has a 0-eigenvalue mode as in Eq. (95), which corresponds to the energy conservation. In a more realistic system, there is always some finite but usually small relaxation rate that comes from the energy being drained from electrons into phonons. Moreover, since the eigenvalue of the mode is especially small, the inverse of the collision operator becomes proportional to projector on the 0-eigenmode. Since the projector reduces the responce to $\partial_T \bar{f}$, the proportionality coefficient can be interpreted as the variation of temperature $\Delta T$ due to the presence of the drive. Thus, the second order correction can be written as

$$f_{0,\lambda=1}^{(2)} \approx \Delta T(t) \frac{\partial \bar{f}}{\partial T}, \tag{106}$$

where the expression for $\Delta T$ is model dependent. For a critical regime numeric inversion of the collision integral suggests a form

$$\Delta T \approx A \frac{k_F^2 v_F^2}{T E_0^2} \sum_{i,j} W_0(\nu_{ij}, T) \tilde{W}_1'(\nu_i, T) \left[ \mathcal{E}_{-\nu_j}^* \mathcal{E}_{\nu_i} + \mathcal{E}_{\nu_j} \mathcal{E}_{-\nu_i}^* \right], \tag{107}$$

where $E_0 = 2k_F^2 v_F/e$. Expression for $W_0$ is the same as in Eq. (104) but $g_0(T, \lambda = 1) \ll g(\omega \sim T, T)$ and in the critical regime

$$A_{\text{cr}} = \frac{3}{4\sqrt{2\pi^2 - 12}}, \qquad \tilde{W}_{k,\text{cr}}'(\nu, T) = \frac{k_F v_F}{i\nu a_{\text{cr}} + \Gamma_k + \tilde{g}_{\text{cr}}'(T)}, \qquad \tilde{g}_{\text{cr}}'(T) = \frac{g'^2 k_F}{4\pi v_F} T \left( \frac{5}{6} + \ln\left( \frac{c_d T}{m^2} \right) \right). \tag{108}$$

Away from the equilibrium, where the Fermi liquid regime takes place, we obtain

$$A_{\text{FL}} = \frac{1}{2\pi}, \qquad \tilde{W}_{1,\text{FL}}'(\nu, T) = \frac{k_F v_F}{i\nu a_{\text{FL}} + \Gamma_1 + \tilde{g}_{\text{FL}}'(T)}, \qquad \tilde{g}'(T) = \frac{g'^2 k_F c_d}{8\pi^2 v_F \Delta^2} \frac{14\pi^2 T^2}{5}. \tag{109}$$

Expression for $A_{\text{cr}}$, $A_{\text{FL}}$, $\tilde{g}_{\text{cr}}'$, and $\tilde{g}_{\text{FL}}'$ is obtained from a numerical inversion of the corresponding linearized collision operators. Even though exact expression might be useful for measuring the relaxation rates, as we will see below, the overall response driven by the effective temperature change can be completely understood through $\Delta T$ only.

The third order perturbation involves the responses attributed to the excitation of the quadrupolar $f_{\pm 2}$ and a density $f_0$ harmonics. We distinguish those by the rates that those involve: the quadrupolar-induced responses would always involve $W_2$, while the density-induced responses will always involve $W_0$. We construct and treat those responses separately. First we focus on the third order response arising from the quadrupolar excitation of the second order.

The third order perturbation of $f_{\pm 1}^{(3)}$ that is associated with a perturbation of a quadrupolar harmonic $f_{\pm 2}^{(2)}$ can be written as

$$f_{-1,\text{quad}}^{(3)} = (f_{+1,\text{quad}}^{(3)})^* = -\frac{e^3}{8k_F^3}\sum_{i,j,l}\mathcal{E}_{\nu_i}\mathcal{E}_{\nu_j}\mathcal{E}_{-\nu_l}^* W_1(\nu_{ijl},\omega,T)\frac{\partial}{\partial\omega}\left[W_2(\nu_{ij},\omega,T)\frac{\partial}{\partial\omega}\left[W_1(\nu_i,\omega,T)\frac{\partial\bar{f}}{\partial\omega}\right]\right]\qquad(110)$$

and is the only term of that type.

The contributions from the perturbations of the density harmonics at the 2nd order of perturbation theory takes a more complicated form. The third order approximation is the leading order at which non-linear corrections appear, in particular, correction to the scattering rate $\delta g[f_0, f_{B0}]$. However, as a non-linear correction, it only plays a role of an extra source term and thus can be easily included in the dynamics. As $\delta g$ correction comes from the density harmonic distortion $f_0^{(2)}$, and thus it has to be attributed to all the the other responses originating from $f_0^{(2)}$. As an additional complication, the details of the third order perturbation become dependent on the value of $\lambda$, since the structure $\delta f_0^{(2)}$ depends on $\lambda$. One $f_0^{(2)}$-driven contribution into $f_{\pm 1}^{(3)}$ comes from the right hand side term in Eq. (92), which we denote as $f_{\pm 1,\text{lin}}^{(3)}$. Another contribution that appears at the 3rd order is coming from the perturbation $\delta g$, we denote it as $f_{\pm 1,\delta g}^{(3)}$. The corresponding expressions take a form

$$f_{-1,\text{lin}}^{(3)} = (f_{+1,\text{lin}}^{(3)})^* = -\frac{e}{2k_F}\sum_i W_1(\nu_{ijl},\omega,T)\mathcal{E}_{\nu_i}\partial_\omega f_0^{(2)}(\omega),\qquad(111)$$

$$f_{+1,\delta g}^{(3)} = (f_{-1,\delta g}^{(3)})^* = -\frac{1}{k_F v_F}\sum_{ijl} W_1(\nu_{ijl},\omega,T)\delta g[f_0^{(2)}]f_1^{(1)}.\qquad(112)$$

where the expressions for $f_0^{(2)}$ and $\delta g$ are $\lambda$-dependent. The expressions for $f_0^{(2)}$ for both $\lambda = 0$ and $\lambda = 1$ have already been shown in Eqs. (105) and (106). Expressions for $\delta g$ differ in the case of $\lambda = 0$ and $\lambda = 1$, because in the $\lambda = 0$ case the bosonic dynamics is absent and $f_{B0}^{(2)} = 0$. The corresponding expressions are

$$\delta g_{\lambda=0} = -\frac{2g'^2 k_F}{v_F}\int_{-\infty}^{+\infty}\frac{d\Omega}{2\pi}B(\Omega)f_{0,\lambda=0}^{(2)}(\omega - \Omega)\qquad(113)$$

$$\delta g_{\lambda=1} = -\frac{2g'^2 k_F}{v_F}\int_{-\infty}^{+\infty}\frac{d\Omega}{2\pi}B(\Omega)\left[f_{0,\lambda=1}^{(2)}(\omega - \Omega) - \frac{1}{\Omega}\int_{-\infty}^{+\infty}d\omega(1 - \bar{f}(\omega + \Omega) - \bar{f}(\omega - \Omega))f_{0,\lambda=1}^{(2)}(\omega)\right]\qquad(114)$$

In case of $\lambda = 1$ a general solution can be constructed in a simple manner due to $f_0^{(2)} \sim \partial_T\bar{f}$. This property simplifies the expression for $\delta g$ in a general form to be $\delta g = \Delta T \partial_T \bar{g}(\omega, T)$. The total response then can be expressed as

$$f_{-1,J}^{(3)} = f_{+1,\text{lin}}^{(3)} + f_{+1,\delta g}^{(3)} = \sum_{ijl}\Delta T(t)\frac{\partial}{\partial T}\left[\left(-\frac{e}{2k_F}\right)W_1(\nu_{ijl},\omega,T)\mathcal{E}_l\partial_\omega\bar{f}(\omega)\right]\qquad(115)$$

The expression for $\lambda = 0$ density response can be written as

$$f_{-1,J}^{(3)} = -\frac{g'^2 k_F}{4\pi v_F}\left(\frac{e}{2k_F}\right)^3\sum_{ijl}\mathcal{E}_l(\mathcal{E}_{-j}^*\mathcal{E}_i + \mathcal{E}_j\mathcal{E}_{-i}^*)\left[W_0(\nu_{ij},T)W_1(\nu_i,\omega,T)W_1(\nu_l,\omega,T)W_1(\nu_{ijl},\omega,T)\left(\frac{\partial\bar{f}}{\partial\omega}\right)^2 + \right.$$
$$\left. + W_1(\nu_{ijl},\omega,T)W_0(\nu_{ij},T)\frac{\partial^2}{\partial\omega^2}\left(W_1(\nu_i,\omega,T)\frac{\partial\bar{f}(\omega)}{\partial\omega}\right)\right]\qquad(116)$$

## B. Current and conductivity

Using the expressions for the perturbations $f^{(1)}$, $f^{(2)}$, and $f^{(3)}$ to compute linear conductivity and corrections to it using expression in Eq. (43) for current. Corresponding linear response $\sigma(\nu, T)$ is obtained by substituting Eq. (102) into Eq. (39) and can be written as

$$\sigma(\nu, T) = \frac{e^2}{4\pi\hbar}\tilde{W}_1(\nu, T),\qquad(117)$$

where we introduce a convenient for the theory notation

$$\tilde{W}_k(\nu, T) = -\int_{-\infty}^{+\infty} d\omega \, W_k(\nu, \omega, T) \, \partial_\omega \bar{f}. \tag{118}$$

For finite $\Gamma_k$ and low enough temperature $T$ and external frequency $\nu$ we can evaluate the integral by expanding $W_k$ as a series in $\bar{g}(\omega, T)$ around $\nu = 0$ and $T = 0$ to obtain

$$\tilde{W}_k(\nu, T) \approx \frac{k_F \, v_F}{i\nu \, a(T) + \Gamma_k + \tilde{g}(T)}, \qquad \tilde{g}(T) = -\int_{-\infty}^{+\infty} d\omega \, g(\omega, T) \, \partial_\omega \bar{f}. \tag{119}$$

where corresponding critical and non-critical expressions take a form

$$\tilde{g}_{\mathrm{cr}}(T) = \alpha_1 T \tilde{\gamma}(T), \qquad \alpha_1 = \frac{g'^2 k_F}{4\pi v_F}, \qquad \tilde{\gamma}(T) = \ln\left(\frac{e}{\pi} \ln\left(\frac{\Lambda_q^2 c^2}{c_d T}\right)\right). \tag{120}$$

$$\tilde{g}_{\mathrm{FL}}(T) = \alpha_2 T^2, \qquad \alpha_2 = \frac{4\pi^2}{3} \frac{g'^2 k_F c_d}{8\pi^2 v_F \Delta^2}, \tag{121}$$

$$\beta_{\mathrm{cr}}(T) = 1 + \frac{\alpha_1}{2\pi} \ln\left(\frac{c^2 \Lambda_q^2}{2\pi c_d T}\right), \tag{122}$$

$$\beta_{\mathrm{FL}} = 1 + \frac{\alpha_1}{2\pi} \ln\left(\frac{c^2 \Lambda_q^2}{e \Delta^2}\right). \tag{123}$$

The values $\alpha_1$, $\alpha_2$, and $a_{\mathrm{FL}}$ are independent of temperature, functions $\tilde{\gamma}(T)$ and $a_{\mathrm{cr}}(T)$ are slowly varying functions of temperature, which can be regarded as roughly constant in a wide range of temperature intervals. Near the critical point we reproduce the conductivity of a strange metal, and away from the criticality we reproduce a Fermi liquid.

The second order response to the current is 0 due to inversion symmetry of the system - there are no excitations to $m = \pm 1$ harmonics in the second in electric field order. The corresponding third order responses lead to a few contributions into the non-linear conductivity. Non-linear contribution into current that arises from the density harmonic excitation in the second order response for $\lambda = 1$ is

$$\mathbf{j}_{\lambda=1}^{(3)} = \Delta T \frac{\partial \sigma}{\partial T} \mathbf{E}. \tag{124}$$

This contribution is nothing more than a change of a current due to a change of resistance with temperature. The analogous response for $\lambda = 0$ cannot be expressed in such a simple way and cannot be interpreted as a response from the temperature change, since the perturbation $f_0^{(2)}$ is non-thermal according to Eq. (105). Thus the perturbation coming from a density harmonic is

$$\mathbf{j}_{J,\lambda=0}^{(3)} = -\frac{e^2}{4\pi E_0^2} \left(\frac{\alpha_1 k_F v_F}{6T}\right) \sum_{ijl} \mathbf{E}_l (\mathbf{E}_{-j}^* \cdot \mathbf{E}_i) \, W_0(\nu_{ij}, T) \tilde{W}_1'(\nu_i, T) \tilde{W}_1'(\nu_l, T) \tilde{W}_1'(\nu_{ijl}, T). \tag{125}$$

The expression above still has a polarization structure similar to a joule heating $\mathbf{E}^2 \mathbf{E}$, but can no longer be interpreted in this way, since $f_0^{(2)}$ does not have a thermal profile. Thus, this response will violate the normal resistance change due to Joule heating with respect to its magnitude, but will still have similar response properties. The non-linear conductivity then can be written as

$$\sigma_{J,\lambda=0}^{(3) \, a,bcd}(\nu_b, \nu_c, \nu_d; T) = -\frac{e^2}{4\pi E_0^2} \left(\frac{\alpha_1 k_F v_F}{6T}\right) \delta_{ab}\delta_{cd} W_0(\nu_{cd}, T) \tilde{W}_1'(\nu_d, T) \tilde{W}_1'(\nu_b, T) \tilde{W}_1'(\nu_{bcd}, T). \tag{126}$$

The third order non-linear response that arises from the quadrupolar harmonic in the second order response can be written as

$$\mathbf{j}_Q^{(3)} = -\frac{e^2}{4\pi} \left(\frac{k_F v_F}{E_0^2}\right) \int_{-\infty}^{+\infty} d\omega \sum_{ijl} \frac{\partial^2 W_1}{\partial \omega^2}(\nu_{ijl}, \omega, T) W_2(\nu_{ij}, \omega, T) W_1(\nu_i, \omega, T) \frac{\partial \bar{f}}{\partial \omega} \mathbf{\Delta}^{abc} E_{\nu_i, a} E_{\nu_j, b} E_{-\nu_l, c} \tag{127}$$

where vector $\mathbf{\Delta}^{abc}$ is $\mathbf{\Delta}^{abc} = (\operatorname{Re} v^a v^b v^{c*}, \operatorname{Im} v^a v^b v^{c*})$, $\mathbf{v} = (1, i)$. Values $E_{\nu, a}$ are $x$- and $y$- components of $\mathbf{E}_\nu$.

After simplifying the expression above, in the leading order in temperature close to the critical point we obtain

$$\sigma_{\text{cr},Q}^{(3),a,bcd}(\nu_b, \nu_c, \nu_d; T) = -\frac{e^2}{4\pi E_0^2} \left( \frac{2\alpha_1 k_F v_F}{3T} \right) \delta_{ab} \delta_{cd} \tilde{W}_1'^2(\nu_{bcd}, T) \tilde{W}_2'(\nu_{cd}, T) \tilde{W}_1'(\nu_d, T). \tag{128}$$

On the other hand, in the Fermi liquid regime we obtain

$$\sigma_{\text{FL},Q}^{(3),a,bcd}(\nu_b, \nu_c, \nu_d; T) = -\frac{e^2}{4\pi E_0^2} \left( \frac{3\alpha_2}{2\pi^2} \right) \delta_{ab} \delta_{cd} \tilde{W}_1'^2(\nu_{bcd}, T) \tilde{W}_2'(\nu_{cd}, T) \tilde{W}_1'(\nu_d, T). \tag{129}$$

One can see from Eq. (127) that quadrupolar responses are susceptible to the second derivative of $W$ over $\omega$, and thus are susceptible to the derivatives of electron self-energies $\Sigma_R$. Hence higher order responses are, in general, susceptible to higher order derivatives of self energy. every two extra orders in $\mathbf{E}$ would add two extra derivatives $\partial_\omega$ and two extra factors of $W_m$ when non-linear responses are considered. Since $\Sigma_R = \Sigma_R(\omega/T)$ close to the criticality, in the leading order in $1/T$ every derivative of $\omega$ will results in the extra factor of $1/T$ contributing to the higher order conductivity. Therefore, the scaling suggests

$$\sigma^{(2n+1)} \sim \frac{E^{2n+1}}{T^{2n-1}} \tag{130}$$

for $n \geq 1$. Thus, non-linear current $\mathbf{j}_{\text{non-lin}}(\mathbf{E}, T)$ has to scale as

$$\mathbf{j}_{\text{non-lin}} \sim T^2 F\left( \frac{\mathbf{E}}{T} \right), \tag{131}$$

where $F$ is a sample-dependent function. In contrast, away from the criticality the dependence of the self-energy goes as $\Sigma_R \sim \omega^2/\Delta^2$, and thus the scaling of the form of Eq. (131) will not take place.

## IV. STRONG FIELD TOY MODEL

### A. Analytical considerations

To qualitatively understand the nonlinear transport of strange metallicity in the regime of nonperturbative strong fields and very low temperatures, one can consider the following toy model. Namely, we simplify Eq.(66) by: (1) neglecting the backaction of bosons on fermions (2) take the scattering rates in all angular momentum channels to be the same (including the energy relaxation $m = 0$) and equal to $\Gamma_0 + \alpha_1 T \log \cosh[\omega/2T]$. Note that to correspond better to the model in main text, $\Gamma_0$ should be taken to be temperature-dependent too, i.e. $\Gamma_0 \rightarrow \Gamma_0 + \alpha_2 T$. Since below we analyze the results for an arbitrary value of $\Gamma_0$, we will take $\Gamma_0$ to be independent of $T$ and then discuss the implications of its $T$-dependence.

Orienting $\mathbf{E}$ along the $x$-direction, we get the following kinetic equation:

$$\begin{aligned} ev_F E \cos\theta \, \partial_\omega f + \Gamma(\omega) f &= \Gamma(\omega) \bar{f}, \\ \Gamma(\omega) &= \Gamma_0 + \alpha_1 T \log \cosh[\omega/2T] \end{aligned} \tag{132}$$

where $\bar{f}(\omega, T)$ is the equilibrium distribution function.

Remarkably, the distribution function can be found exactly at any temperature with $\bar{f}(\omega, T) = 1/(e^{\omega/T} + 1)$ being a Fermi distribution at temperature $T$, and the relaxation rate $\Gamma(\omega, T) = \Gamma_0 + \alpha_1 T \log \cosh(\omega/2T)$. Non-perturbative solution over $E$ in this case can be written in a form $f(\omega, \theta, T) = \bar{f}(\omega, T) + \delta f(\omega, \theta, T)$, where

$$\begin{cases} \delta f(\omega, \theta, T) = \exp\left[ -\frac{\int_0^\omega d\omega' \Gamma(\omega')}{ev_F E \cos\theta} \right] \int_{-\infty}^\omega d\omega' \exp\left[ \frac{\int_0^{\omega'} d\omega'' \Gamma(\omega'')}{ev_F E \cos\theta} \right] \left[ -\partial_\omega \bar{f}(\omega', T) \right] & \cos\theta > 0 \\ \delta f(\omega, \theta, T) = \exp\left[ -\frac{\int_0^\omega d\omega' \Gamma(\omega')}{ev_F E \cos\theta} \right] \int_\omega^\infty d\omega' \exp\left[ \frac{\int_0^{\omega'} d\omega'' \Gamma(\omega'')}{ev_F E \cos\theta} \right] \partial_\omega \bar{f}(\omega', T) & \cos\theta < 0 \end{cases} \tag{133}$$

With the help of the distribution function above one can write down the expression for current $j$ as

$$j = \frac{ek_F}{8\pi^2} \int_{-\pi/2}^{\pi/2} d\theta \cos\theta I(\theta),$$

$$I(\theta) = \int_{-\infty}^{\infty} d\omega \exp\left[-\frac{\int_0^{\omega} d\omega' \Gamma(\omega')}{ev_F E \cos\theta}\right] \int_{-\infty}^{\omega} d\omega'' \exp\left[\frac{\int_0^{\omega''} d\omega' \Gamma(\omega')}{ev_F E \cos\theta}\right] \frac{1}{T \cosh^2 \frac{\omega}{2T}} =$$

$$= \int_{-\infty}^{\infty} d\omega \int_0^{\infty} d\delta \exp\left[-\frac{\int_{\omega}^{\omega+\delta} d\omega' \Gamma(\omega')}{ev_F E \cos\theta}\right] \frac{1}{T \cosh^2 \frac{\omega}{2T}} \qquad (134)$$

$$= \int_{-\infty}^{\infty} d\omega \int_0^{\infty} d\delta \exp\left[-\frac{\Gamma_0 \delta + F(\omega, \delta)}{ev_F E \cos\theta}\right] \frac{1}{T \cosh^2 \frac{\omega}{2T}},$$

$$F(\omega, \delta) = \int_{\omega}^{\omega+\delta} d\omega' \alpha_1 T \log \cosh[\omega'/2T]$$

### B. Simple limits

Let us start with discussing the limit of $T = 0$, where $\bar{f}(\omega > 0) = 0$ and $\bar{f}(\omega < 0) = 1$ and $\Gamma(\omega) \to \Gamma_0 + \alpha_1 |\omega|/2$. The solution to this linear equation can be easily constructed, in combination with the boundary conditions $f(\omega \to \infty, \theta) = 0$ and $f(\omega \to -\infty, \theta) = 1$ it leads to

$$\begin{cases} f = 0, & \omega > 0, \cos\theta < 0, \\ f = 1 - \exp\left[-\frac{\int_0^{\omega} d\omega' \Gamma(\omega')}{ev_F E \cos\theta}\right], & \omega < 0, \cos\theta < 0, \end{cases} \qquad \begin{cases} f = \exp\left[-\frac{\int_0^{\omega} d\omega' \Gamma(\omega')}{ev_F E \cos\theta}\right], & \omega > 0, \cos\theta > 0, \\ f = 1, & \omega < 0, \cos\theta > 0, \end{cases} \qquad (135)$$

With the current given by

$$j = ek_F \int \frac{d\omega}{2\pi} \int \frac{d\theta}{2\pi} \cos\theta \, f(\omega, \theta), \qquad (136)$$

the current for $\gamma(\omega) = \Gamma + \alpha_1 |\omega|/2$ becomes

$$j = \frac{e^2 k_F v_F}{4\pi \Gamma} E, \qquad E \ll \frac{\Gamma^2}{\alpha_1 ev_F} \qquad (137)$$

$$j = \frac{3ek_F K(1/2)}{4\pi} \sqrt{\frac{ev_F E}{\pi \alpha_1}}, \qquad E \gg \frac{\Gamma^2}{\alpha_1 ev_F}. \qquad (138)$$

In the first regime - the regime of relatively small $E$, the leading response is linear, which is identical to the expression at $T = 0$ obtained from perturbation theory. The second regime, which is characterized by strong fields $E$, has a scaling of $j \sim \sqrt{E}$, which explicitly reflects the energy dependence of scattering rate.

Note that the nonanalytic $j(E)$ behavior at large $E$ does not necessarily imply a non-Fermi liquid scattering. The calculation for the same toy model in the Fermi liquid regime can be carried out with $\Gamma(\omega) = \Gamma + \alpha_2 \omega^2$, instead of linear in $\omega$ dependence. The corresponding currents are:

$$j = \frac{e^2 k_F v_F}{4\pi \Gamma} E, \qquad E \ll \frac{\Gamma}{ev_F} \sqrt{\frac{\Gamma}{\alpha_2}} \qquad (139)$$

$$j = ek_F \frac{\Gamma\left(\frac{4}{3}\right) \Gamma\left(\frac{1}{6}\right)}{8\pi^{\frac{3}{2}} \Gamma\left(\frac{2}{3}\right)} \left(\frac{3ev_F E}{\alpha_2}\right)^{\frac{1}{3}}. \qquad (140)$$

Note the different power scaling of the current $j \sim E^{1/3}$ in the Fermi liquids in strong fields, which is slower than $j \sim \sqrt{E}$ in strange metals. Since one generally expects $\alpha_1 \sim 1$ in strange metals and $\alpha_2 \sim 1/T_F$ in Fermi liquids, the crossover to the non-linear regime in strange occurs at values of $E$ lower by a factor of $\sqrt{T_F/\Gamma}$ than in Fermi liquids.

Another interesting limit is the limit of high temperatures $\alpha_1 T/\Gamma \gg 1$. In this case there is only one free dimensionless parameter in the system: $E' = ev_F E/\alpha_1 T^2$. The current takes a form

$$j(E) = T f_{T \to \infty} \left(\frac{ev_F E}{\alpha_1 T^2}\right), \qquad (141)$$

| | Electric field regime | | |
|---|---|---|---|
| Temperature regime | $j \propto ek_F v_F$. leading term; subleading term | | |

| | $\tilde{E} \ll \Gamma_0 T$ | $\Gamma_0 T \ll \tilde{E} \ll \frac{\Gamma_0^2}{\alpha_1}$ | $\frac{\Gamma_0^2}{\alpha_1} \ll \tilde{E}$ |
|---|---|---|---|
| $\Gamma_0 \gg T$ | $\frac{E}{\Gamma_0}; \frac{T^2}{\Gamma_0} f_{sc}\left(\frac{E}{\Gamma_0 T}\right)$ | $\frac{E}{\Gamma_0}; \alpha_1 \frac{E^2}{\Gamma_0^2} \operatorname{sgn}[E]$ | $\sqrt{E/\alpha_1}; \frac{\Gamma_0}{\sqrt{\alpha_1}}$ |

| | $\tilde{E} \ll \sqrt{\frac{\Gamma_0^3 T}{\alpha_1}}$ | $\sqrt{\frac{\Gamma_0^3 T}{\alpha_1}} \ll \tilde{E} \ll \alpha_1 T^2$ | $\alpha_1 T^2 \ll \tilde{E}$ |
|---|---|---|---|
| $\Gamma_0 \ll T$ | $\frac{E}{\sqrt{\Gamma_0 \alpha_1 T}}; \frac{T^2}{\Gamma_0} f_{sc}\left(\frac{E}{\Gamma_0 T}\right)$ | $\frac{E^{2/3}}{\sqrt{\alpha_1} T^{1/3}}; \frac{\Gamma_0}{\alpha_1^{5/6}}$ | $\sqrt{E/\alpha_1}; \sqrt{\alpha_1} T$ |

TABLE I. Summary of different regimes of $j(E)$ dependence for the toy model.

where

$$f_{T \to \infty}(E') = -2ek_F \int_{-\frac{\pi}{2}}^{+\frac{\pi}{2}} d\theta \cos\theta \int_{-\infty}^{+\infty} \frac{dx}{2\pi} e^{-\frac{F_\infty(x)}{E' \cos\theta}} \int_{-\infty}^{x} dx' \partial_x \bar{f}(x'T, T) e^{\frac{F_\infty(x')}{E' \cos\theta}}, \tag{142}$$

where

$$F_\infty(x) = 2x \log 2 + \operatorname{Li}_2\left(-e^{-x}\right) - \operatorname{Li}_2\left(-e^x\right). \tag{143}$$

## C. Crossover behavior: leading and subleading contributions to $j(E)$

The discussion below illustrates how to obtain the qualitative form of $j(E)$ in various limits (results are summarized in Table I). The integrals over $\theta$ do not change the scaling with respect to the dimensionful quantities $(E, \Gamma_0, T)$, and therefore its outcome is not discussed in detail. For convenience, we define

$$\tilde{E} \equiv ev_F E, \tag{144}$$

which has the dimensions of $\Gamma_0^2$.

Low temperatures: $\Gamma_0 \gg \alpha_1 T$

• $\tilde{E} \ll \Gamma_0 T$: In the limit of low $E$ and $T$, the integral over $\delta$ is dominated by $\exp\left[-\frac{\Gamma_0 \delta}{ev_F E \cos\theta}\right]$, leading to a characteristic value of $\delta \sim \frac{ev_F E \cos\theta}{\Gamma_0}$. This allows to use the expansion of $F(\omega, \delta)$ in $\delta$:

$$F(\omega, \delta)_{\delta \ll \omega} \approx \delta\alpha_1 T \log\cosh[\omega/2T] + \alpha_1 \tanh[\omega/2T]\frac{\delta^2}{4} + \frac{\alpha_1 \delta^3}{24 T \cosh^2[\omega/2T]} + O(\delta^4). \tag{145}$$

In the limit $\Gamma_0 \gg \alpha_1 T$ The first term contributes an overall prefactor that is of the order $1 - O(\alpha_1 T/\Gamma)$. We can then use perturbation theory to compute corrections, expanding $\exp\left[-\frac{F[\omega, \delta]}{ev_F E \cos\theta}\right]$. The $\propto \delta^2$ term correction vanishes after the integral over $\omega$. The lowest-order correction then comes from the $\propto \delta^3$ and leads to a contribution $I_{nonlin}^{(3)} \propto \frac{ev_F E \cos\theta}{\Gamma_0} \frac{\alpha_1}{T ev_F E \cos\theta}\left(\frac{ev_F E \cos\theta}{\Gamma_0}\right)^3$ such that $j_{nonlin}^{(3)} \propto \frac{E^3}{\Gamma_0^4 T}$, while the linear part of the current is $j_{lin} \propto \frac{E}{\Gamma_0}$. To go beyond the perturbative expansion, one notices that $F(\omega, \delta) = T^2 \phi\left(\frac{\delta}{T}, \frac{\omega}{2T}\right)$. As long as $T^2 \phi\left(\frac{\delta}{T}, \frac{\omega}{2T}\right) \ll \Gamma_0 \delta$, we can use $e^{-F} \approx 1 - F$. This results in the nonlinear correction to current (after variable change $\omega/2T \to x$, $\delta\Gamma_0/(ev_F E) \to y$) being a universal function:

$$j_{nonlin} = \frac{ek_F}{4\pi^2} \int_{-\pi/2}^{\pi/2} d\theta \int_{-\infty}^{\infty} dx \int_{0}^{\infty} dy \frac{T^2}{\Gamma_0} \phi\left(\frac{ev_F E}{\Gamma_0 T} y, x\right) \frac{e^{-\frac{y}{\cos\theta}}}{\cosh^2 x} = \frac{T^2}{\Gamma_0} f_{sc}\left(\frac{E}{\Gamma_0 T}\right) \tag{146}$$

.

• $\Gamma_0 T \ll \tilde{E} \ll \frac{\Gamma_0^2}{\alpha_1}$: In this regime, the convergence of the $\delta$ integral is still controlled by $\Gamma_0$, i.e. $\Gamma_0 \delta \gg F(\omega, \delta)$. However, the characteristic $\delta$ is $\sim \frac{E}{\Gamma_0} \gg T$ in this case. One can then approximate $\frac{1}{T \cosh^2 \frac{\omega}{2T}}$ with a delta function and obtain $F(\omega \to 0, \delta) \approx \frac{\alpha_1 \delta^2}{4}$ using $\delta \gg \omega$. The remaining integral has the form:

$$I(\theta) \approx 4 \int_{0}^{\infty} d\delta \exp\left[-\frac{\Gamma_0 \delta + \alpha_1 \delta^2/4}{ev_F E \cos\theta}\right]. \tag{147}$$

In principle, this is the result one obtains in the $T \to 0$ limit, discussed above. The leading nonlinear correction to the current in $E \ll \frac{\Gamma_0^2}{\alpha_1}$ limit is then found to be $j_{nonlin} \propto \alpha_1 \frac{E^2}{\Gamma_0^2}$. This form is not inconsistent with presence of inversion symmetry; for a different sign of $E$ one can show that $j_{nonlin}$ will also change sign. Therefore $j_{nonlin} \propto \alpha_1 \frac{E^2}{\Gamma_0^2} \operatorname{sgn}[E]$. We note that this result corresponds to the scaling form, Eq. (146), in the limit $f_{sc}(x \to \infty)$. Therefore, $f_{sc}(x \to \infty) \propto x^2$.

- $\frac{\Gamma_0^2}{\alpha_1} \ll \tilde{E}$: Here Eq. (147) holds too, but the second term in the exponential is dominant. The leading term for current in this case is $j \propto \sqrt{E/\alpha_1}$, while correction due to finite $\Gamma_0$ is of the order $\delta j \propto \frac{\Gamma_0}{\sqrt{\alpha_1}}$ (note that it is independent of $E$)

<u>High temperatures: $\Gamma_0 \ll \alpha_1 T$</u> Note that this limit may not necessarily be practically achievable since $\Gamma_0$ can be temperature dependent too. In that case, the condition for this limit to hold is $\Gamma_0(T) \ll \alpha_1 T$.

- $\tilde{E} \ll \sqrt{\frac{\Gamma_0^3 T}{\alpha_1}}$: In this regime we expect characteristic $\delta$ to be small, specifically $\delta \ll T$, which allows to use the Taylor expansion of $F(\omega, \delta)$ in $\delta$. The only difference with $\Gamma_0 \gg \alpha_1 T$ is that we can no longer neglect the effect of linear term in $\delta$. Specifically, in lowest order in $\delta$ we need to solve

$$I(\theta) \approx \int_{-\infty}^{\infty} d\omega \int_0^{\infty} d\delta \exp\left[-\frac{\Gamma_0 \delta + \delta \alpha_1 T \log \cosh[\omega/2T]}{e v_F E \cos \theta}\right] \frac{1}{T \cosh^2 \frac{\omega}{2T}}. \qquad (148)$$

The integral over $\omega$ can be carried out explicitly using $\int_{-\infty}^{\infty} \frac{dx}{\cosh^{2+a} x} \frac{\sqrt{\pi} \Gamma\left(\frac{a}{2}+1\right)}{\Gamma\left(\frac{a+3}{2}\right)} \approx_{a \gg 1} \frac{2.5}{\sqrt{a}}$ ($a = \frac{\delta \alpha_1 T}{e v_F E \cos \theta}$ here). The integral over $\delta$ then takes the form $\int_0^{\infty} d\delta \frac{\sqrt{e v_F E \cos \theta}}{\sqrt{\delta}} e^{-\frac{\Gamma_0 \delta}{E}}$, where we used the limit $a \gg 1$ which is correct for $\delta \sim \frac{E}{\Gamma_0}$. One notices that for $\Gamma_0 = 0$ the integral diverges; in this case $\Gamma_0$ thus provides a convergence factor, such that the result is $j_{lin} \propto \frac{E}{\sqrt{\Gamma_0 \alpha_1 T}}$. Note that this is not inconsistent with linear in $T$ resistivity - as mentioned above $\Gamma_0$ in the model in main text is actually equal to $\Gamma_0 \to \Gamma_0 + \alpha_2 T$ and thus for $\alpha_2 T \gg \Gamma_0$ one gets $\rho \propto T$. The nonlinear correction in this regime exhibits the same scaling behavior as for $\Gamma_0 \gg \alpha_1 T$ due to the same arguments. To obtain the upper limit of $E$ for this regime, we consider the next regime first.

- $\sqrt{\frac{\Gamma_0^3 T}{\alpha_1}} \ll \tilde{E} \ll \alpha_1 T^2$: We define this regime as the one where $\Gamma_0$ can be neglected. We thus enter the $\Gamma_0 = 0$ regime discussed above; the scaling regime (146) thus ends here. The convergence of the integral over $\delta$ should then be ensured by the $\delta^3$ term ($\delta^2$ term on its own does not make the integral convergent). The cutoff for $\delta$ is then set by $\frac{\alpha_1 \delta^3}{TE}$ becoming of the order 1, i. e. $\delta_{cut} \sim (ET/\alpha_1)^{1/3}$. For the current we thus obtain $j \propto \sqrt{\delta_{cut} E/(\alpha_1 T)} \propto E^{2/3}/(\sqrt{\alpha_1} T^{1/3})$. The correction from finite $\Gamma_0$ is $\propto \Gamma_0/\alpha_1^{5/6}$

This result allows us to define the crossover field between the two regions discussed above: it occurs when the cutoff $\delta_{cut}$ becomes of the order $E/\Gamma_0$ (the cutoff provided by $\Gamma_0$), i.e. for $E \propto \sqrt{\frac{\Gamma_0^3 T}{\alpha_1}}$. The upper bound of the present regime occurs when $\delta \sim T$, corresponding to $E \sim \alpha_1 T^2$

- $\alpha_1 T^2 \ll \tilde{E}$: Here we can approximate the integral over $\omega$ with that of a delta function and set $\omega \to 0$ in $F(\omega \to 0, \delta)$. This leads to the same result as in $\Gamma_0 \gg \alpha_1 T$ case, namely $j \propto \sqrt{E/\alpha_1}$. The correction due to finite $\Gamma_0$ is also the same; however, because $\Gamma_0 \ll \alpha_1 T$, the corrections due to finite $T$ are more relevant in this case. The expansion of $F(0, \delta)$ for $\delta \to \infty$ takes the form

$$F(0, \delta \gg T) \approx \alpha_1 \frac{\delta^2}{4} - \alpha_1 T \delta \log[2] + O(T^2). \qquad (149)$$

Expanding the exponential, one can evaluate the correction to the result. One gets:

$$\frac{\delta j}{j_{E \to \infty}} = \frac{\int_{-\pi/2}^{\pi/2} d\theta \int_0^{\infty} d\delta \frac{\alpha_1 T \delta \log[2]}{e v_F E} \exp\left[-\frac{\delta^2}{4 e v_F E \cos \theta}\right]}{\int_0^{\pi/2} d\theta \cos \theta \int_0^{\infty} d\delta \exp\left[-\frac{\delta^2}{4 e v_F E \cos \theta}\right]} = \frac{6 \log 2}{\sqrt{2\pi} K(1/2)} \sqrt{\frac{\alpha_1 T^2}{E}} \approx 0.89 \sqrt{\frac{\alpha_1 T^2}{E}}. \qquad (150)$$

The resulting correction to current thus scales as $\delta j \propto \sqrt{\alpha_1} T$, i.e. independent of $E$.

### D. Numerical results

We now demonstrate the actual extent of the scaling regime at low $E$ with a numerical integration of Eq. (134). For numerical integration, we perform a variable change $\omega \to A_0 \frac{a+b}{\sqrt{2}}$; $\omega' \to A_0 \frac{a+b}{\sqrt{2}}$ and normalize all quantities by a

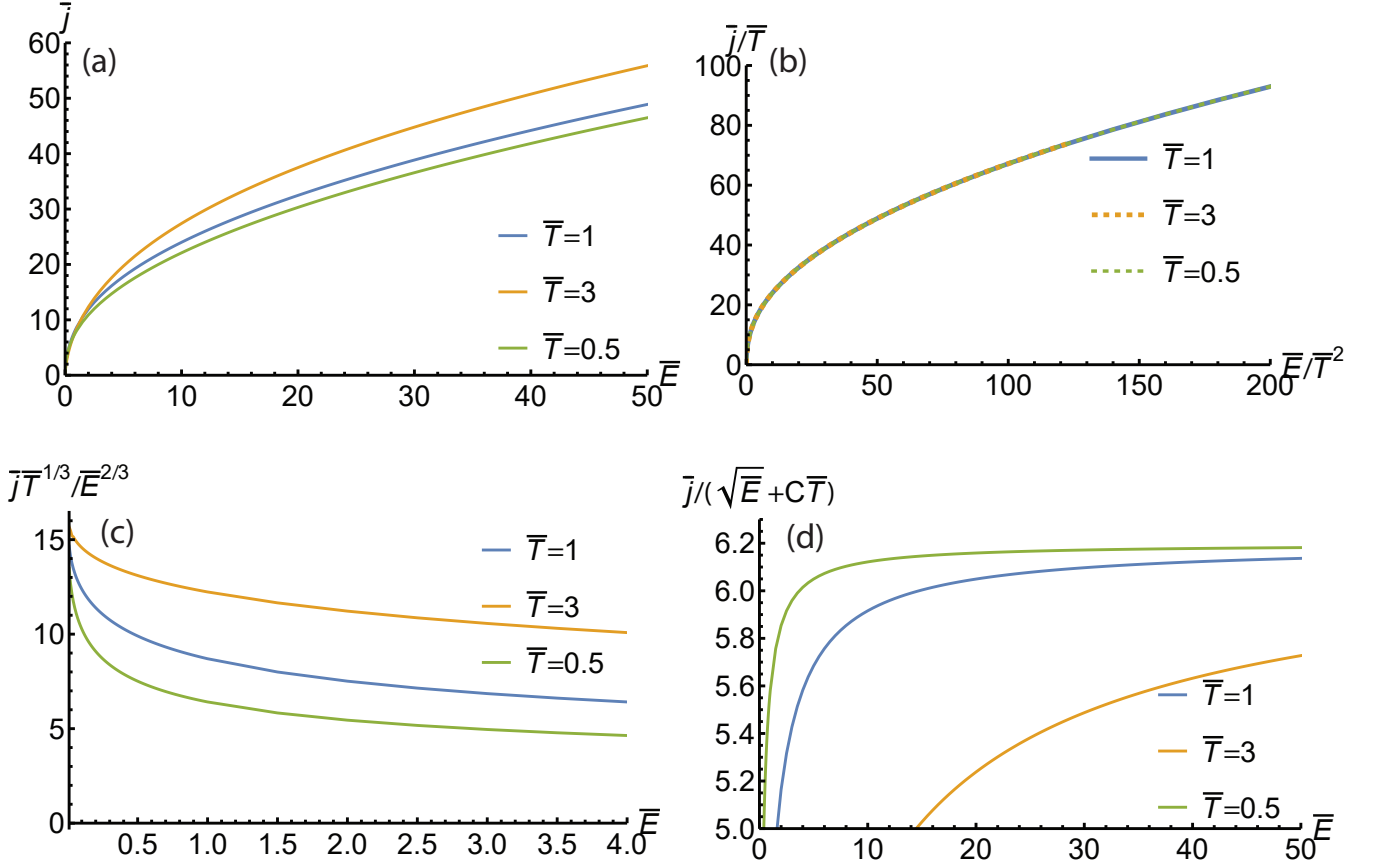

FIG. 1. Properties of the toy model for $\overline{\Gamma}_0 = 0$, $\alpha_1 = 1$ (a) Current-voltage $\overline{j}(\overline{E})$ for different values of $\overline{T}$. (b) Scaled plot $\overline{j}(\overline{E}/\overline{T}^2)$; curves with different $\overline{T}$ overlap. (c) $\overline{j}/\overline{T}^{1/3}\overline{E}^{2/3}$ tends to the same constant value at low $\overline{E}$, confirming the low-$E$ behavior $\overline{j} \propto \overline{E}^{2/3}/\overline{T}^{1/3}$ in Tab. I. (d) $\overline{j}/(\sqrt{\overline{E}} + C\overline{T})$ ($C \approx 0.89$ from Eq. (150)) tends to the same constant value at high $\overline{E}$, confirming the high-$E$ behavior of $j$ in Tab. I.

single energy scale $A_0$. The resulting expression evaluated numerically is

$$\overline{j} = \frac{4\pi^2 A_0}{ek_F}j = \int_{-\frac{\pi}{2}}^{\frac{\pi}{2}} d\theta \cos\theta \int_{-\infty}^{\infty} da \int_0^{\infty} db \frac{\exp\left\{-G[\frac{a+b}{\sqrt{2}}, \overline{T}] + G[\frac{a-b}{\sqrt{2}}, \overline{T}]\right\}}{2\overline{T}\cosh^2\frac{a-b}{2\sqrt{2}\overline{T}}}$$

$$G(x, \overline{T}) = \overline{\Gamma}_0 x + \alpha_1\left(\overline{T}\left(\overline{T}\mathrm{Li}_2\left(-e^{-\frac{x}{\overline{T}}}\right) - \frac{x^2}{4\overline{T}} - x\log\left(e^{-\frac{x}{\overline{T}}} + 1\right) + x\log\left(\cosh\left(\frac{x}{2\overline{T}}\right)\right)\right)\right) + \frac{\pi^2\overline{T}^2}{12}$$

$(151)$

where $\overline{E} = \frac{ev_F E}{A_0^2}$, $\overline{T} = \frac{T}{A_0}$, $\overline{\Gamma}_0 = \frac{\Gamma_0}{A_0}$.

The results for $\Gamma_0 = 0$ are presented in Fig. 1. One observes that curves at different $\overline{T}$ collapse on a universal curve when scaled against $\overline{E}/\overline{T}^2$. Moreover, the predictions (Tab. I) are also confirmed for the low- (c) and high- (d) $\overline{E}$ limits.

In Fig. 2 we present the numerical results for $\overline{\Gamma}_0 = 1$. First of all, the cubic nonlinear conductivity (c) shows a strong growth at lower temperatures. The strong dependence of $\overline{\sigma}$ on $\overline{T}$ deviates from the low-T limit and is not reaching the high-T limit discussed in (Tab. I). Thus we have attempted to use the scaling $\frac{\overline{T}^2}{\overline{\Gamma}_0}f_{sc}\left(\frac{\overline{E}}{\overline{\Gamma}_0\overline{T}}\right)$ by empirically replacing $\Gamma_0$ with $1/\sigma$. The results (e) show, that a good degree of convergence can be obtained for a range of temperatures and $\overline{E} < 0.25\overline{T}$.

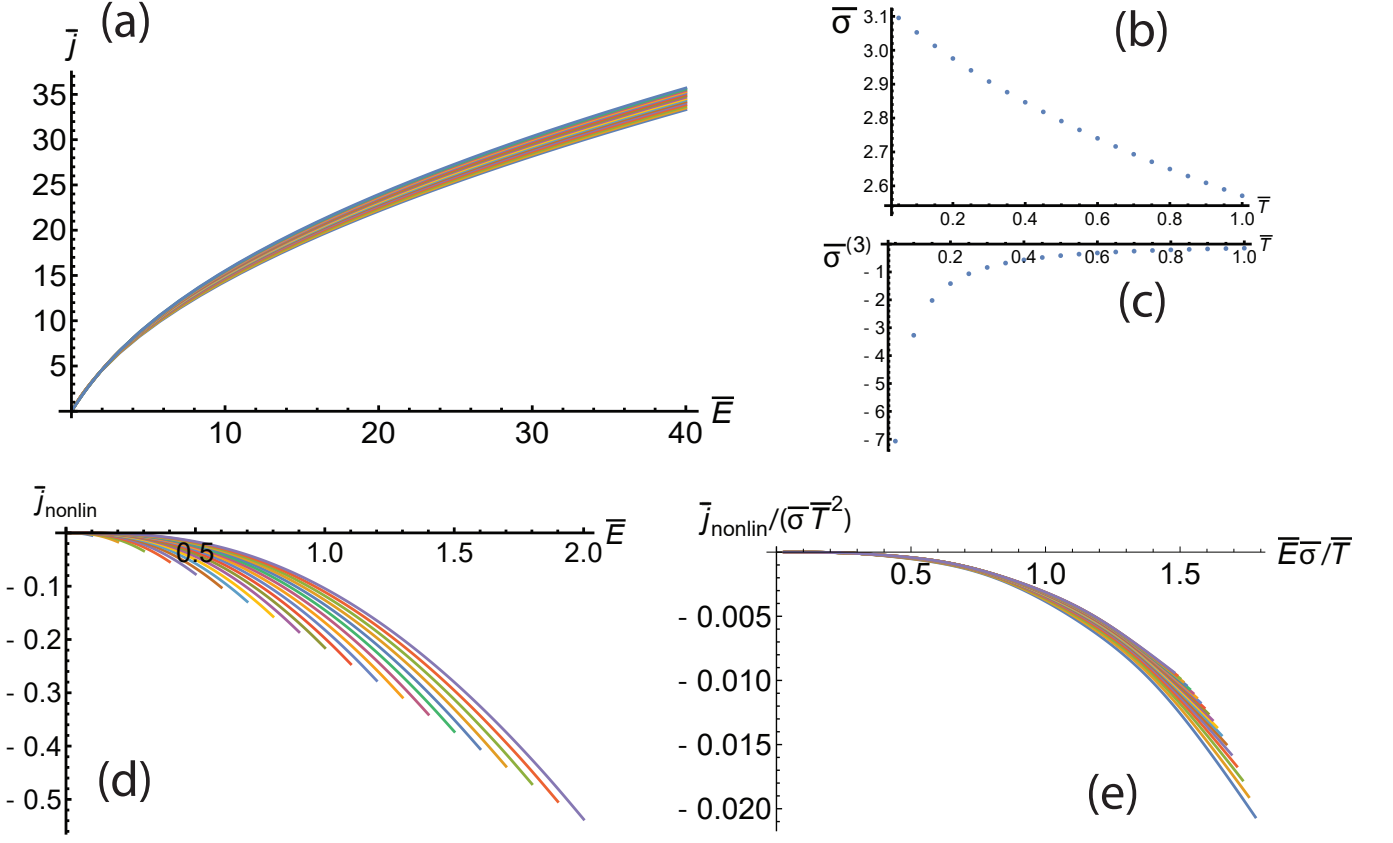

FIG. 2. Properties of the toy model for $\overline{\Gamma}_0 = 1$, $\alpha_1 = 1$ (a) Current-voltage $\overline{j}(\overline{E})$ for different values of $\overline{T}$ ranging from 0.25 to 1. (b) Linear and (c) cubic conductivity extracted from (a) at $\overline{E} < 0.1\overline{T}$ as a function of temperature. (d) nonlinear current obtained by subtracting $\overline{\sigma}\overline{E}$ from $j$ in (a) for $\overline{T} \in [0.05, 1]$ (e) Scaling plot for $j$ with $\overline{E} < 0.25\overline{T}$ (cf. Tab. I, here we empirically extended the low$-T$ limi t by $1/\Gamma_0 \to \sigma$).

## V. THIRD-ORDER RESPONSE FOR LATTICE SYSTEMS

Here we propose and analyze a heuristic generalization of the kinetic equations above to lattice systems with discrete rotations symmetry. To work out the responses for a $D_{4h}$ symmetric system we expand the responses into irreducible representations of $D_{4h}$ instead of expanding the response in angular harmonics. We assume that scattering rates are fixed for each separate irreducible representation. Thus, instead of using $W_m(\nu, \omega, T)$, where $m$ is a number of an angular momentum, one should use $W_\alpha$, where $\alpha \in \{A_{1g}, E_u, B_{1g}, B_{2g}\}$ - irreducible representations of $D_{4h}$. Thus, for example, $W_{E_u} = k_F v_F/(ia_{E_u}(T) + \Gamma_{E_u} + g_{E_u}(\omega, T))$. We will assume a form of $g_\alpha(\omega, T)$ and $a_\alpha(T)$ similar to the spherically symmetric $g(\omega, T)$ and $a(T)$ dependence on $T$ and $\omega$.

By applying considerations analogous to a circularly symmetric case, one can construct the third-order response mediated by quadrupolar excitations of the distribution of $B_{1g}$ and $B_{2g}$ irreducible representations as

$$\mathbf{j}_{Q,B_{1g}/B_{2g}}^{(3)} = -A_{Q,B_{1g}/B_{2g}} \frac{e^2}{4\pi\hbar E_0^2} \sum_{ijl} \mathbf{\Phi}_{B_{1g}/B_{2g}}(\mathbf{E}, \nu_l, \nu_j, \nu_i) \tilde{W}_{E_u}'^2(\nu_{ijl}) \tilde{W}_{B1/B2}'(\nu_{ij}) \tilde{W}_{E_u}'(\nu_i) \tag{152}$$

Where for $B_{1g}$ we have

$$\mathbf{\Phi}_{B_{1g}}(\mathbf{E}, \nu_l, \nu_j, \nu_i) = \mathbf{e}_x E_{\nu_l,x}(E_{\nu_j,x}E_{\nu_i,x} - E_{\nu_j,y}E_{\nu_i,y}) + \mathbf{e}_y E_{\nu_l,y}(E_{\nu_j,y}E_{\nu_i,y} - E_{\nu_j,x}E_{\nu_i,x}), \tag{153}$$

and for $B_{2g}$ we have

$$\mathbf{\Phi}_{B_{2g}}(\mathbf{E}, \nu_l, \nu_j, \nu_i) = \mathbf{e}_x E_{\nu_l,y}(E_{\nu_j,y}E_{\nu_i,x} + E_{\nu_j,x}E_{\nu_i,y}) + \mathbf{e}_y E_{\nu_l,x}(E_{\nu_j,y}E_{\nu_i,x} + E_{\nu_j,x}E_{\nu_i,y}). \tag{154}$$

Corresponding expressions for $A_{Q,B_{1g}/B_{2g}}$ are

$$A_{Q,B_{1g}} = A_Q \int_0^{2\pi} \frac{d\theta}{2\pi}(\hat{v}_{F,x}^4(\theta) - \hat{v}_{F,x}^2(\theta)\hat{v}_{F,y}^2(\theta)), \qquad A_{Q,B_{2g}} = A_Q \int_0^{2\pi} \frac{d\theta}{2\pi}\hat{v}_{F,x}^2(\theta)\hat{v}_{F,y}^2(\theta). \tag{155}$$

Vector $\hat{v}_F = (\hat{v}_{F,x}, \hat{v}_{F,y})$ is a unit vector in the direction of Fermi velocity. When all $\mathbf{E}_{\nu_i}$ are directed either along $x$ or $y$, only $\mathbf{j}_{Q,B_{1g}}^{(3)}$ is non-zero and $\mathbf{j}_{Q,B_{2g}}^{(3)} = 0$. When all $\mathbf{E}_{\nu_i}$ are directed along the diagonal in the $xy$ plane, $\mathbf{j}_{Q,B_{1g}}^{(3)} = 0$ and $\mathbf{j}_{Q,B_{2g}}^{(3)}$ is non-zero.

Since these responses have different magnitude due to $\Gamma_{B_{1g}} \neq \Gamma_{B_{2g}}$, when applied electric field is not directed along $x$, $y$, or $xy$ diagonal, the non-linear current arising from quadrupolar responses $\mathbf{j}_{Q,B_{1g}}^{(3)} + \mathbf{j}_{Q,B_{2g}}^{(3)}$ will not be collinear with $\mathbf{E}$. Moreover, the perpendicular to $\mathbf{E}$ part of the current will consists only from the quadrupolar responses, since the response arising from density $A_{1g}$ representation will always be collinear with $\mathbf{E}$. Thus, measuring the perpendicular to $\mathbf{E}$ component of a total non-linear current $\mathbf{j}^{(3)}$ allows direct access to the quadrupolar responses.

## VI. ESTIMATES FOR BI-2212

We chose Bi-2212, since experimental information both about transport and Fermi surface properties of this material is available. First, we estimate the non-linear correction to conductivity from quadrupolar processes, not arising from heat. For this we assume that all the decay rates $\Gamma_{m \neq 0} = \Gamma$ are equal and that $k_F \sim E_F/V_F$. This allows us to estimate the non-linear correction $\sigma_{\pm 2}^{(3)}$ from the value of linear sheet conductivity $\sigma$.

Sections 3 and 13 of Supplemental Information of the work [7] contain the information about bulk resistivity and single conducting sheet thickness, which allows us to obtain relevant sheet resistivity data. For temperature taken at $T \sim 40$ K: $\rho \sim 40\,\mu\Omega \cdot$ cm (see Section 3 Fig. 3 d of Supplemental information of [7] for $T = 20$ K), $T_F \sim 2 \cdot 10^3 K$, single sheet thickness $d \sim 0.8$ nm (see Section 13, Table S2 of [7]), $v_F \sim 3 \cdot 10^5$ m/s [8, 9]. Cited data and assumptions above are sufficient to estimate the orders of magnitude of of $\bar{W}_1(\nu = 0, T = 40\,\text{K}) \sim 100$ with use of Eq. (117), $E_0 = 2k_F^2 v_F/e \sim 1$MV/cm and $A_Q \sim T_F/T \sim 50$, which results in the value for the non-linear conductivity $\sigma^{(3)} \sim 3 \cdot 10^{-4} A_Q\, \Omega^{-1} \text{m}^{-1} (\text{V/m})^{-2}$.

For pulsed field experiments, we can also estimate nonlinear response due to heating; for a pulse of $\sim 10$ ps[10] and magnitude of $E = 50$ V/cm, specific heat $c \sim 50$ mJ/gK [11], mass density $\rho_m \sim 6.5 \cdot 10^6$ g/m$^3$ [12], and $d\rho/dT \sim 0.6\,\mu\Omega$ cm/$K$ (see Section 13, Table S2 of [7]), the temperature raise per pulse will be $\Delta T \sim 2$ mK, which will result in non-linear correction to conductivity $\sigma_J^{(3)} \sim 3 \cdot 10^{-6}\,\Omega^{-1}\text{m}^{-1}(V/m)^{-2}$ and relative value $\sigma_J^{(3)}/\sigma^{(1)} = \frac{\Delta \tau}{c \rho_m \rho}\frac{d\rho/dT}{\rho} \approx 10^{-12}(V/m)^{-2}$.

One can also imagine a transport experiment at extremely low temperature. For $T \sim 10K$ obtained with the methods similar to [7] and [13] and electric pulse of $\tau \sim 0.1\,\mu s$, one would expect a similar order of magnitude $\sigma_{\pm 2}^{(3)}$ response to the one computed above, and $\sigma_0^{(3)} \sim 1 \cdot 10^{-2}\,\Omega^{-1}\text{m}^{-1}(V/m)^{-2}$, which for $A_Q \sim 200$ still results in the non-heating related component being slightly stronger.



\end{document}